\documentstyle[epsfig,amsfonts,amsbsy]{elsart}
\let\sectiontmp\section\let\subsectiontmp\subsection \let\appendixtmp\appendix
\def\section{\setcounter{equation}{0}\sectiontmp}
\def\subsection{\subsectiontmp}
\def\theequation{\arabic{section}.\arabic{equation}}
\def\appendix{\def\theequation{\Alph{section}.\arabic{equation}}\appendixtmp}
\def\propagatorR#1#2#3#4{\begin{picture}(1.4,.8)#3
    \if#4r
    \put(1.3,.1){\circle{.2}}\put(1.1,.3){$R$}\else
    \put(.1,.1){\circle{.2}}\put(-0.1,.3){$R$}\fi
    \put(1.3,0.1){\vector(-1,0){.8}}\put(.1,0.1){\line(1,0){.6}}
    \put(.2,-.15){\makebox(0,0){#1}}
    \put(1.2,-.15){\makebox(0,0){#2}}\end{picture}}
\def\propagatorA#1#2#3#4{\begin{picture}(1.4,.6)#3
    \if#4r
    \put(1.3,.1){\circle{.2}}\put(1.1,.3){$R$}\else
    \put(.1,.1){\circle{.2}}\put(-0.1,.3){$R$}\fi
    \put(0.1,0.1){\vector(1,0){.8}}\put(.7,0.1){\line(1,0){.6}}
    \put(.2,-.15){\makebox(0,0){#1}}
    \put(1.2,-.15){\makebox(0,0){#2}}\end{picture}}
\def\self#1#2#3{\begin{picture}(2.2,.8)\thicklines
    \put(2.3,0.1){\vector(-1,0){1.3}}\put(-.1,0.1){\line(1,0){1.3}}
    \put(1.1,0.1){\oval(2,1)}
    \put(0,-0.15){\makebox(0,0){#1}}\put(2.2,-0.15){\makebox(0,0){#2}}
    \if#3R {\put(0.1,0.1){\circle{.2}}}\else
    \if#3A {\put(2.1,0.1){\circle{.2}}}\else
    {\put(1.1,0.1){\makebox(0,0){#3}}}
    \fi\fi\end{picture}}

\def\contourxy{\unitlength=0.8cm
\begin{picture}(17,1)\thicklines
\contour
\put(14.5,-.5){\makebox(0,0){$\infty$}}
\put(11,-.5){\makebox(0,0){$t_x^+$}}
\put(8,1.8){\makebox(0,0){$t_y^-$}}
\put(11,0){\circle*{.2}}
\put(8,1){\circle*{.2}}
\end{picture}}

\def\contour{\thicklines
\put(1,-.5){\makebox(0,0){$t_{0}$}}
\put(16.5,.5){\makebox(0,0){$t$}}
\put(0,.5){\vector(1,0){16}}
\put(1,1.){\line(1,0){13}}\put(14,.5){\oval(1,1)[br]}
\put(14,0){\vector(-1,0){13}}\put(14,.5){\oval(1,1)[tr]}}

\linethickness{1.0pt}
\def\geight{\begin{picture}(20,0)
\put(0,2){
     \thicklines
     \put(10,0){\circle*{2.00}}
     \put(10,7.5){\circle{15.05}}
     \put(10,-7.5){\circle{15.05}}}
     \end{picture}}
\def\sloop{\begin{picture}(20,0)
\put(0,2){
     \thicklines
     \put(10,0){\circle*{4.00}}
     \put(10,7.5){\circle{15.00}}}
     \end{picture}}
\def\GGJ{\begin{picture}(20,15)
\put(0,2){
     \thicklines
     \put(10,0){\circle*{2.00}}
     \put(10,7.5){\circle{15.00}}
     \put(10,7.5){\makebox(0,0){$\ii J$}}
     \put(10,0){\thinlines\line(0,-1){10}}}
     \end{picture}}
\def\GPb{\begin{picture}(10,0)
\put(0,2){\put(5,3){\thinlines\line(0,-1){6}}
          \put(5,3){\makebox(0,0){$\times$}}}
     \end{picture}}
\def\GPbfull{\begin{picture}(10,0)
\put(0,2){\put(5,3){\line(0,-
1){6}}\put(5,3){\makebox(0,0){$\bigotimes$}}}
     \end{picture}}
\def\GG{\begin{picture}(24,0)
\put(22,2){\thinlines\vector(-
1,0){12}}\put(2,2){\thinlines\line(1,0){10}}
     \end{picture}}
\def\GGfull{\begin{picture}(24,0)
\put(22,2){\thicklines\vector(-
1,0){12}}\put(2,2){\thicklines\line(1,0){10}}
     \end{picture}}
\def\GGxy{\begin{picture}(24,0)
\put(22,2){\thinlines\vector(-
1,0){12}}\put(2,2){\thinlines\line(1,0){10}}
\put(4,-3){\makebox(0,0){$x$}}\put(20,-3){\makebox(0,0){$y$}}
     \end{picture}}
\def\GGfullxy{\begin{picture}(24,0)
\put(22,2){\thicklines\vector(-
1,0){12}}\put(2,2){\thicklines\line(1,0){10}}
\put(4,-3){\makebox(0,0){$x$}}\put(20,-3){\makebox(0,0){$y$}}
     \end{picture}}
\def\Dysonself{\begin{picture}(60,0)
\put(0,2){\thicklines
     \put(0,0){\thinlines\line(1,0){10}}
        \put(20,0){\thinlines\vector(-1,0){12}}
     \put(60,0){\vector(-1,0){12}}\put(40,0){\line(1,0){10}}
     \put(30,0){\oval(20,15)}
     \put(30,0){\makebox(0,0){$-\ii\Se$}}}
     \end{picture}}
\def\gloopphitwo{\begin{picture}(20,0)
\put(0,2){
     \thicklines
     \put(10,0){\line(-1,-1){6}}
     \put(10,0){\line(1,-1){6}}
     \put(10,0){\circle*{2.00}}
     \put(4,-6){\makebox(0,0){$\bigoplus$}}
     \put(16,-6){\makebox(0,0){$\bigoplus$}}
     \put(10,7.5){\circle{15.00}}}
     \end{picture}}
\def\gloopphione{\begin{picture}(20,0)
\put(0,2){
     \thicklines
     \put(10,0){\line(0,-1){6}}
     \put(10,0){\circle*{4.00}}
     \put(10,-6){\makebox(0,0){$\bigotimes$}}
     \put(10,7.5){\circle{15.00}}}
     \end{picture}}
\def\gphifour{\begin{picture}(22,0)
\put(4,2){
     \thicklines
     \put(2,5){\line(1,-1){10}}
     \put(12,5){\line(-1,-1){10}}
     \put(7,0){\circle*{2.00}}
     \put(2,-5){\makebox(0,0){$\bigoplus$}}
     \put(12,-5){\makebox(0,0){$\bigoplus$}}
     \put(2,5){\makebox(0,0){$\bigoplus$}}
     \put(12,5){\makebox(0,0){$\bigoplus$}}}
     \end{picture}}
\def\sphitwo{\begin{picture}(22,0)
\put(4,2){
     \thicklines
     \put(7,0){\line(1,1){6}}
     \put(4.5,0){\line(1,0){5}}
     \put(7,0){\line(-1,1){6}}
     \put(7,0){\circle*{4.00}}
     \put(2,5){\makebox(0,0){$\bigoplus$}}
     \put(12,5){\makebox(0,0){$\bigoplus$}}}
     \end{picture}}
\def\gphithree{\begin{picture}(22,0)
\put(4,2){
     \thicklines
     \put(2,0){\line(1,0){12}}
     \put(8,0){\line(0,1){6}}
     \put(8,0){\circle*{4.00}}
     \put(14,0){\makebox(0,0){$\bigotimes$}}
     \put(8,6){\makebox(0,0){$\bigotimes$}}
     \put(2,0){\makebox(0,0){$\bigotimes$}}}
     \end{picture}}
\def\gsand4{\begin{picture}(38,0)
\put(2,2){
     \thicklines
     \put(2,0){\circle*{2.00}}
     \put(32,0){\circle*{2.00}}
     \bezier{216}(2,0)(17,-10)(32,0)
     \bezier{216}(2,0)(17,+10)(32,0)
     \bezier{284}(2,0)(17,-25)(32,0)
     \bezier{284}(2,0)(17,+25)(32,0)}
     \end{picture}}
\def\ssand3{\begin{picture}(38,0)
\put(2,2){
     \thicklines
     \put(2,0){\line(1,0){30}}
     \put(2,0){\circle*{4.00}}
     \put(32,0){\circle*{4.00}}
     \bezier{284}(2,0)(17,-25)(32,0)
     \bezier{284}(2,0)(17,+25)(32,0)}
     \end{picture}}
\def\sphisand2phi{\begin{picture}(42,0)
\put(2,2){
     \thicklines
     \put(2,0){
     \put(2,0){\line(1,0){8}}
     \put(24,0){\line(1,0){8}}
     \put(10,0){\makebox(0,0){$\bigotimes$}}
     \put(24,0){\makebox(0,0){$\bigotimes$}}
     \put(2,0){\circle*{4.00}}
     \put(32,0){\circle*{4.00}}
     \bezier{284}(2,0)(16,-25)(32,0)
     \bezier{284}(2,0)(16,+25)(32,0)}}
     \end{picture}}
\def\gphisand3phi{\begin{picture}(53,0)
\put(2,2){
     \thicklines
     \put(3,0){\line(1,0){44}}
     \put(3,0){\makebox(0,0){$\bigotimes$}}
     \put(47,0){\makebox(0,0){$\bigotimes$}}
     \put(8,0){
     \put(2,0){\circle*{2.00}}
     \put(32,0){\circle*{2.00}}
     \bezier{284}(2,0)(16,-25)(32,0)
     \bezier{284}(2,0)(16,+25)(32,0)}}
     \end{picture}}
\def\gloopphih{\begin{picture}(43,0)
\put(2,2){
     \thicklines
     \put(2,0){\line(1,0){37}}
     \put(39,0){\makebox(0,0){$\bigotimes$}}
     \put(2,0){\circle*{4.00}}
     \put(32,0){\circle*{2.00}}
     \bezier{284}(2,0)(16,-25)(32,0)
     \bezier{284}(2,0)(16,+25)(32,0)}
     \end{picture}}
\def\DSa{\begin{picture}(0,0)\thicklines
\put(0,0){\oval(1.5,1)}
\put(0,0){\makebox(0,0){$-\ii\Sa$}}\end{picture}}
\def\GlnG0Sa{
\begin{picture}(4.3,1.5)
\put(0.5,.1){
\put(.5,0){\DSa}\put(3,0){\DSa}\put(1.75,1){\DSa}
\put(1.75,-1){\makebox(0,0){\dots\dots}}
\put(1,.5){\oval(1,1)[lt]}\put(2.5,.5){\oval(1,1)[rt]}
\put(1,-.5){\oval(1,1)[lb]}\put(2.5,-.5){\oval(1,1)[rb]}}
\end{picture}}
\def\GGaSa{
\begin{picture}(2.5,2.)
\thicklines\put(0.5,.1){
\put(.5,0){\DSa}\put(2,-.5){\line(0,1){1}}
\put(1.,1){\line(1,0){0.5}}\put(1.,-1){\line(1,0){0.5}}
\put(1,.5){\oval(1,1)[lt]}\put(1.5,.5){\oval(1,1)[rt]}
\put(1,-.5){\oval(1,1)[lb]}\put(1.5,-.5){\oval(1,1)[rb]}}
\end{picture}}
\def\vhight#1{\vphantom{\left(\begin{picture}(0,#1)\end{picture}\right)}
}
\def\Dclosed#1#2{
\begin{picture}(2,.8)\put(0,.1){#2
\put(1,0){\circle{1.414}}
\put(1,-.707){\line(0,1){1.414}}\put(.293,0){\line(1,0){1.414}}
\put(0.5,-0.5){\line(0,1){1}}\put(1.5,-0.5){\line(0,1){1}}
\put(0.5,-0.5){\line(1,0){1}}\put(0.5,0.5){\line(1,0){1}}}
\put(1.5,-.8){$#1$}
\end{picture}}
\def\Dclosedone#1#2{
\begin{picture}(2,.8)\put(0,.1){#2
\put(1,0){\circle{1.414}}
\put(1,-.707){\line(0,1){1.414}}\put(.293,0){\line(1,0){1.414}}
\put(0.5,-0.5){\line(0,1){1}}\put(1.5,-0.5){\line(0,1){1}}
\put(0.5,-0.5){\line(1,0){1}}\put(0.5,0.5){\line(1,0){1}}}
\put(1.5,-.8){$#1$}
\put(.293,0.1){\circle*{.2}}
\end{picture}}
\newcommand{\di}{{\mathrm d}}
\newcommand{\Tr}{{\mathrm{Tr}}}
\newcommand{\ii}{{\mathrm i}}

\renewcommand{\and}{\quad{\mathrm{and}}\quad}

\renewcommand{\Re}{{\mathrm{Re}}}
\renewcommand{\oint}{\int_{\cal C}}

\def\Tc{{\cal T}_{\cal C}}

\def\scr#1{\mbox{\scriptsize #1}}

\def\vec#1{\mbox{\boldmath $#1$}}
\newcommand{\dpi}[1]{\frac{\di^4 #1}{(2\pi)^4}}                
                    
\newlength{\charwidth}
\def\medhat#1{\settowidth{\charwidth}{$#1\,$}{\makebox[\charwidth]{$\,
 {\widehat{\makebox[2mm]{$#1\,$}}}$}}\vphantom{#1}}
\newcommand{\lap}%
{\raisebox{-0.5ex}{$\stackrel{\scriptstyle <}{\scriptstyle \sim}$}}
\newcommand{\gap}%
{\raisebox{-0.5ex}{$\stackrel{\scriptstyle >}{\scriptstyle \sim}$}}
 
\def\Gr{\Delta}\def\Se{\Pi}
\def\j{\medhat{J}}
\def\ja{\medhat{J}}\def\jad{\medhat{J}^\dagger}

\def\ea{\eta}\def\Ka{K}
\def\Pa{\medhat{\phi}}\def\Pad{\Pa^\dagger}
\def\Pta{\medhat{\varphi}}\def\Ptad{\Pta^\dagger}
\def\Pt{\medhat{\varphi}}\def\Ptd{\Pt^\dagger}
\def\PtI{\medhat{\varphi}_{\rm I}}
\def\Ph{\medhat\phi}\def\Phd{\medhat{\phi}^{\dagger}}

\def\Pba{{\phi}}\def\jba{{J}}
\def\dc{\delta_{\cal C}}
\def\Ga{\Gr}
\def\Sa{\Se}
\def\Sb{\Se}                                   
\def\A{A}                                   
\def\Gm{\Gamma}                                   
\def\suma{\displaystyle \left(\frac{1}{2}\right)_{\scr{neut.}}}
\def\Lg{{\cal L}}
\def\Lgh{\makebox[3.5mm]{${\widehat{\makebox[2mm]{$\Lg$}}}$}\vphantom{L}}
\def\Lint{\Lgh^{\mbox{\scriptsize int}}}\def\LintI{\Lint_{\rm I}}

\def\Hh{\medhat{H}}

\def\So{S} 
 
\makeatletter
\def\ps@copyright{\let\@mkboth\@gobbletwo
  \def\@oddhead{}%
  \let\@evenhead\@oddhead
  \def\@oddfoot{\small\sl
      GSI-Preprint-98-34, subm. to Nucl. Phys. B
	\hfill }
  \let\@evenfoot\@oddfoot
}
\begin{document}
\begin{frontmatter}
\title{Self-Consistent Approximations to\\ Non-Equilibrium Many-Body Theory} 

\author{Yu. B. Ivanov$^{1,2}$, J. Knoll$^{1}$ 
and D. N. Voskresensky$^{1,3}$} 

\maketitle

\noindent                        
$^{1}${\it\small Gesellschaft f\"ur Schwerionenforschung mbH, Planckstr. 1,
64291 Darmstadt, Germany}
\\
$^{2}${\it\small Kurchatov Institute, Kurchatov sq. 1, Moscow 123182,
Russia} 
\\
$^{3}${\it\small Moscow Institute for Physics and Engineering, 
Kashirskoe sh. 31, Moscow 115409, Russia}                    

\begin{abstract}

Within the non-equilibrium Green's function technique on the real time
contour, the $\Phi$-functional method of Baym is reviewed and generalized to
arbitrary non-equilibrium many-particle systems. The scheme may be closed at
any desired order in the number of loops or vertices of the generating
functional. It defines effective theories, which provide a closed set of
coupled classical field and Dyson equations, which are self-consistent,
conserving and thermodynamically consistent.  The approach permits to include
unstable particles and therefore unifies the description of resonances with
all other particles, which obtain a mass width by collisions, decays or
creation processes in dense matter. The inclusion of classical fields enables
the treatment of soft modes and phase instabilities. The method can be taken
as a starting point for adequate and consistent quantum improvements of the
in-medium rates in transport theories.
\end{abstract}
\end{frontmatter}

\section{Introduction}

Non-equilibrium Green's function technique, developed by Schwinger, Kadanoff,
Baym and Keldysh \cite{Schw,Kad62,Keld64,Lif81}, is the appropriate concept to
study the space--time evolution of many-particle quantum systems. This
formalism finds now applications in various fields, such as quantum
chromodynamics \cite{Land}, nuclear physics
\cite{Dan84,Dan90,Toh,Bot90,MSTV,Vos93,Knoll95}, astrophysics
\cite{MSTV,VS87,Keil}, cosmology \cite{CalHu}, spin systems \cite{Manson},
lasers \cite{Korenman}, physics of plasma \cite{Bez,Kraft}, physics of liquid
$^{3}$He \cite{SerRai}, critical phenomena, quenched random systems and
disordered systems \cite{Chou}, normal metals and super-conductors
\cite{VS87,Rammer,Fauser}, semiconductors \cite{LipS}, tunneling and secondary
emission \cite{Noziers}, etc.

For actual calculations certain approximation steps are necessary.  In many
cases perturbative approaches are insufficient, as for physical systems with
strong couplings, e.g. like those treated in nuclear physics, in physics of
liquids $^{3}$He and $^{4}$He, or high-temperature super-conductivity, etc.
In such cases, one has to re-sum certain sub-series of diagrams in order to
obtain a reasonable approximation scheme. In contrast to perturbation theory
for such re-summations one frequently encounters the complication that the
resulting equations of motion, even though self-consistent, may no longer
comply with the conservation laws, e.g., of charge, energy and momentum.  This
problem has first been considered in two pioneering papers by Baym and
Kadanoff \cite{KadB,Baym} discussing the response to an external perturbation
of quantum systems in thermodynamic equilibrium. Baym, in particular, showed
\cite{Baym} that any approximation, in order to be conserving, must be
so-called $\Phi$-derivable.  Thereby, he exploited the properties of an
auxiliary functional, the $\Phi$-functional, introduced by Luttinger and Ward
\cite{Luttinger} a year earlier for the formulation of the thermodynamic
potential (see also \cite{Abrikos}).  Thereby the $\Phi$-functional is
determined in terms of full, i.e. re-summed, Green's functions and free
vertices. The scaling parameter of the vertices can be considered as an
expansion parameter of a given approximation level. In the non-equilibrium
formalism the problem of conserving approximations could be even more severe
than in the case of the systems response to an external perturbation close to
thermal equilibrium, since the system may exercise a rather violent
evolution. Apart from transport models, mostly based on the quasi-particle
approximation like Landau's Fermi liquid theory, there were only few attempts
to discuss the issue of conserving approximations in the context of the
non-equilibrium field theory (see, e.g., \cite{Kad62,Dan84,Bot90}), which
mainly considered Hatree-Fock and T-matrix approximations. However, the
general problem of constructing conserving approximations in the
non-equilibrium case and, in particular, beyond the quasi-particle limit has
not explicitly been addressed yet.

Alongside, the question of thermodynamic consistency is vital. If, as a result
of a non-equilibrium evolution, a system arrives at an equilibrium state, the
non-equilibrium Green's functions should properly describe thermodynamic
quantities and potentials, such that thermodynamic relations between them are
preserved. This problem is also relevant to the thermodynamic Green's function
technique, as already considered by Baym \cite{Baym}. Baym demonstrated that
any $\Phi$-derivable approximation is at the same time thermodynamically
consistent.

In this paper we re-address the above problems and extend the concept i) to
the genuine non-equilibrium case formulated on the closed real-time contour,
ii) to relativistic field-theory Lagrangians in principle of arbitrary type
(not just two-body fermion interactions) and iii) to the inclusion of
classical fields, i.e. non-vanishing expectation values of the field
operators. The generalized scheme permits to construct self-consistent,
approximate, coupled dynamical equations of motion for the classical fields
and Green's functions of the system on the closed real-time contour. This set
of equations is conserving and thermodynamically consistent.  Thereby, the
inclusion of classical fields permits to account for the phase-transition
phenomena or to describe the coherent dynamics of soft modes, much in the
spirit of hard-thermal-loop re-summations \cite{Blaizot,Jackiw93,Knoll95}.
Avoiding the quasi-particle limit enables us to appropriately consider the
finite mass-width of all constituents in the dense matter environment. The
latter aspect unifies the description of resonances, which have already a
mass-width in vacuum, with all particles, which acquire a dynamical width
during the collision processes in the dense matter. The proper account of the
finite width of the particles is also vital for a non-singular treatment of
the soft-mode problem \cite{Knoll95}, where space-time coherence effects, like
the Landau--Pomeranchuk--Migdal effect, defer the use of zero width
quasi-particles and require non-perturbative re-summations. In this paper we
confine the presentation to the derivation of the closed set of
self-consistent couple Kadanoff-Baym and classical field equations. This
constitutes the basis for various further steps towards classical-type
transport schemes through the gradient approximation, which will be presented
in a forthcoming paper\cite{IKV}.

For the sake of clarity, we restrict the presentation to systems of
relativistic scalar bosons. This allows us to formulate the basic ideas in a
simple and transparent form. The resulting relations can directly be
generalized to multi-component systems of relativistic bosons and fermions. In
sect.  2 we introduce the general equations of motion and the expressions for
the conserved quantities on the operator level.  The equations of motion of
the corresponding expectation values are formulated within the real-time
closed contour formalism (sect. 3). Thereby, it is advantage to formulate the
concepts in terms of generating functionals on the non-equilibrium contour,
where the special functional $\Phi$ plays a central role (sect.  4). This
generating functional takes the same status in the space of Green's functions
(two-point functions) and classical fields (one-point functions), as the
original Lagrangian for the field operators. Subsequently, we formulate the
diagrammatic representation for $\Phi$ (sect.  5). We show that any
approximation, where all classical field sources and self-energies are
$\Phi$-derivable in the sense of a variational principle, has the following
properties: (i) it is conserving, (ii) it provides conserved current and
energy-momentum tensor, which are identical to the corresponding Noether
quantities (sect. 6), and (iii) it is at the same time thermodynamically
consistent (sect. 7). In the summary, we formulate the main results and
briefly discuss extensions and applications of the derived formalism.  The
list of diagrammatic rules is deferred to the Appendix A, while Appendix B
contains some helpful equilibrium relations.

\section{Energy-Momentum Tensor and Conserved Currents}
\label{sect-Prel}

We consider a system of relativistic scalar bosons, specified by the free
Klein-Gordon Lagrangians
\begin{eqnarray}
\Lgh^0 =
\left\{
\begin{array}{ll}
\frac{1}{2}\displaystyle\vphantom{\frac{1}{m}}
 \left(\partial_\mu\Pa \cdot \partial^\mu\Pa 
-  m^2 \Pa^2\right)
\quad&{\mbox{for neutral bosons}},\\[2mm] 
\displaystyle
\partial_\mu\Pad\cdot\partial^\mu\Pa
- m^2 \Pad\Pa 
\quad&{\mbox{for charged bosons}},
\end{array}\right.
\label{L0cb}
\end{eqnarray}
%
where $\Pa(x)$ and $\Pad(x)$ are bosonic field operators.  The convention of
units is such that $\hbar=c=1$. The interaction Lagrangians $\Lint\{\Pa\}$
(for neutral bosons) and $\Lint\{\Pa,\Pad\}$ (for charged bosons) are assumed
to be local, i.e. without derivative coupling. Under these conditions the
Lagrangians are charge symmetric. Charges are understood as the
electric charge, strangeness, iso-spin, etc.

The variational principle of stationary action leads to the Euler--Lagrange
equations of motion for the field operators
%
\begin{eqnarray}
\label{eqmotion}
-\So_x \Pa(x)&=& \ja(x)=\frac{\partial
\Lint}{\partial\Pad},\quad\quad\mbox{where }\quad 
\So_x=
-\partial_\mu\partial^\mu -m^2 ,
\end{eqnarray}
%
and similarly for the corresponding adjoint equation. Thereby, the $\ja(x)$
operator is a local source current of the field $\Pa$, while $\So_x$ is the
differential operator of the free evolution with the free propagator
$\Ga^0(y,x)$ as resolvent.

The standard canonical energy-momentum tensor \cite{Itz80} has some undesired
features, as it is non-symmetric in the Lorentz indices, for example.
Alternatively, using the Euler--Lagrange equations of motion and the
definition of the source current (\ref{eqmotion}), one can show that the
following form also defines a conserving energy momentum tensor
%
\begin{eqnarray}
\label{E-M-new-tensor}
\nonumber
\medhat{\Theta}^{\mu\nu}_{\scr{(alt.)}}(x)&=&
-\frac{1}{2}\left[
\suma
\left(\partial_x^\nu-\partial_y^\nu\right)
\left(
\frac{\partial \Lgh^{0}(x)}{\partial\left(\partial_\mu\Pa\right)}
\Pa(y)
-
\Pad(x) 
\frac{\partial \Lgh^{0}(y)}{\partial\left(\partial_\mu\Pad\right)}
\right)
\right]_{x=y}
\nonumber
\\
&&+
g^{\mu\nu}
\left(
{\widehat{\cal E}}^{\scr{int}}(x)-
{\widehat{\cal E}}^{\scr{pot}}(x)
\right) \\
&&\mbox{with} \quad  
	\partial_{\mu}\medhat{\Theta}^{\mu\nu}_{\scr{(alt.)}}(x)=0.
\nonumber
\end{eqnarray}
%
For notational simplicity, expression (\ref{E-M-new-tensor}) and similar
expressions below, which appear symmetric in $\Pa$ and $\Pad$, are written in
such a way that they directly apply to complex fields. The symbol
$(1/2)_{\scr{neut.}}$ implies that for real fields the corresponding
expressions are obtained from those for complex fields by multiplying by the
factor $1/2$ upon equating $\Pad=\Pa$. Above, we have introduced operators of
the interaction energy density ${\widehat{\cal E}}^{\scr{int}}(x)$ of the
system, which accounts for the total interaction part of the energy density,
and the potential energy density ${\widehat{\cal E}}^{\scr{pot}}(x)$, both
given as
%
\begin{eqnarray}
\label{eps-pot}
{\widehat{\cal E}}^{\scr{int}}(x) &=& - \Lint(x),
\nonumber\\
{\widehat{\cal E}}^{\scr{pot}}(x)
&=&-\frac{1}{2}\suma
\left(
\frac{\partial
\Lint}{\partial\Pa} \Pa(x) + \Pad(x) \frac{\partial
\Lint}{\partial\Pad}
\right)\\
&=& - \frac{1}{2}\suma
\left(
\jad(x) \Pa(x) +  \Pad(x) \ja(x)
\right). \nonumber
\end{eqnarray}
%
For a multi-component system, the latter defines the sum of the
potential energy densities of any field 
$\Pa(x)$ with the currents $\ja(x)$ induced by the other fields in the system.

In terms of the differential operator $\widehat{p}^{\mu}_x = \ii
\partial^\mu_x$, the alternative energy--momentum tensor of eq.
(\ref{E-M-new-tensor}) can be written in a charge symmetric form as
%
\begin{eqnarray}
\label{E-M-pi-tensor}
\nonumber
\widehat{\Theta}^{\mu\nu}_{\scr{(alt.)}}(x)
&=&
\frac{1}{4}\suma
\left[
\left(\left(\widehat{p}^{\nu}_x\right)^* 
+ \widehat{p}^{\nu}_y\right)
\left(\left(\widehat{p}^{\mu}_x\right)^* 
+ \widehat{p}^{\mu}_y\right)
\left(
\Pad(x)\Pa(y)
+\Pa(y)\Pad(x)\right)
\right]_{x=y}\\ 
&+&
g^{\mu\nu}
\left({\widehat{\cal E}}^{\scr{int}}(x)-
{\widehat{\cal E}}^{\scr{pot}}(x)
\right).
\end{eqnarray}
%
This form of the energy--momentum
tensor is equal to the metric one, which results from variation of the
action over the metric tensor $g_{\mu\nu}$ rather than over the fields. All
terms in eq. (\ref{E-M-pi-tensor}) are evidently symmetric in $\mu\nu$.

For specific interactions, eq. (\ref{eps-pot})
provides simple relations between ${\widehat{\cal E}}^{\scr{int}}(x)$ and
${\widehat{\cal E}}^{\scr{pot}}(x)$. If all vertices of $\Lint$ have the same
number $\gamma$ of field operators attached, one simply deduces

%
\begin{eqnarray}\label{gamma}
{\widehat{\cal E}}}^{\scr{int}}(x)
=\frac {2}{\gamma}{\widehat{\cal E}^{\scr{pot}}(x). 
\end{eqnarray}
%
For instance, for the $\phi^4$-theory, where $\gamma=4$, the
interaction energy is half of the potential energy.

If the Lagrangian is invariant under some global transformation of charged
fields (with the charge $e$), e.g.,
%
\begin{equation}
\label{c-global-tr.} 
\Pa(x)\Rightarrow e^{-\ii e\Lambda}\Pa(x);\quad\quad
\Pad(x)\Rightarrow e^{\ii e\Lambda}\Pad(x), 
\end{equation}
%
there exists a Noether current defined as  
\cite{Itz80} 
%
\begin{equation}
\label{c-current} 
\medhat{j}^{\mu}_{\scr{(Noether)}}
=  
\displaystyle \left.
\frac{\partial \Lgh(\Lambda(x),\partial_\mu\Lambda(x))}%
{\partial\left(\partial_\mu\Lambda(x)\right)}\right|_{\Lambda(x)=0}, 
\end{equation}
%
which is conserved, i.e. $\partial_{\mu} \medhat{j}^{\mu}_{\scr{(Noether)}} =
0.$ Formally, it is derived by applying the local transformation of the form
(\ref{c-global-tr.}) and using the stationary condition of the action
around physical solutions.

 From the Euler--Lagrange equations of motion (\ref{eqmotion}),
one obtains 
%
\begin{equation}
\label{c-current-ndiv} 
\partial_{\mu} \medhat{j}^{\mu}_{\scr{(Noether)}}=
-\ii e \left(\jad(x)\Pa(x)-\Pad (x)\ja(x)\right), 
\end{equation}
%
which vanishes, since for the symmetry (\ref{c-global-tr.}) the interaction
Lagrangian has to consist of terms, where $\Pa(x)$ and $\Pad(x)$ appear
pairwise.  In terms of the $\widehat{p}^{\mu}_x = \ii \partial^\mu_x$
operator, the $\medhat{j}^{\mu}_{\scr{(Noether)}}$ current can again be
written in a charge symmetric form as
%
\begin{equation}
\label{c-pi-current} 
\medhat{j}^{\mu}_{\scr{(Noether)}}(x) 
= e 
\frac{1}{2}
\left[
\left(\left(\widehat{p}^{\mu}_x\right)^* 
+ \widehat{p}^{\mu}_y\right)
\left(
\Pad(x)\Pa(y)
+\Pa(y)\Pad(x)\right)
\right]_{x=y},
\end{equation}
%
which naturally vanishes for the neutral particles ($e=0$).

One may also define the tensor $M^{\mu\nu\rho}$, which is associated with the
Lorentz invariance of the Lagrangian and provides the angular momentum
conservation. However, we do not treat this tensor in this paper, since it is
of no common use in kinetics.

\section{Real-Time Contours}

In the non-equilibrium case, one assumes that the system has been prepared at
some initial time $t_0$ described in terms of a given density operator
$\medhat{\rho}_0=\sum_a P_a\left|a\right>\left<a\right|$, where the
$\left|a\right>$ form a complete set of eigenstates of $\medhat{\rho}_0$. All
observables can be expressed through $n$-point Wightman functions of
Heisenberg operators $\medhat{A}(t_1),\dots ,\medhat{O}(t_n)$ at some later
times
%
\begin{eqnarray}\label{corrfct}
\left<\medhat{O}(t_n)\right. \left.\dots \medhat{B}(t_2)\medhat{A}(t_1)
\right>
&=:&\Tr\; \medhat{O}(t_n)\dots \medhat{B}(t_2)\medhat{A}(t_1)
\medhat{\rho}_0(t_0)
\nonumber\\
&=&
\sum_a P_a\left<a\right|
\medhat{O}(t_n) \dots \medhat{B}(t_2) \medhat{A}(t_1)
\left|a\right>.  
\end{eqnarray}
%
Note the fixed operator ordering for Wightman functions.

\parbox[t]{14.5cm}{
\begin{center}\vspace*{1cm}
\contourxy\\[1cm]
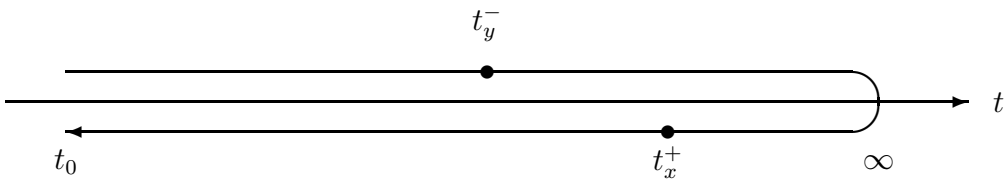
Figure 1: Closed real-time contour with two external points $x,y$ on the
contour. 
\end{center}}

The non-equilibrium theory can entirely be formulated on a {\em special}
contour---the {\em closed real-time contour} (see figure 1) with the time
argument running from $t_{0}$ to $\infty$ along the {\em time-ordered} branch
and back to $t_{0}$ along the {\em anti-time-ordered} branch. Contour-ordered
multi-point functions are defined as expectation values of contour ordered
products of operators
%
\begin{equation}\label{cont.exp.}
\left<\Tc \medhat{A}(x_1)\medhat{B}(x_2)\dots\right>
=\left<\Tc \medhat{A}_{\rm I}(x_1)\medhat{B}_{\rm I}(x_2)\dots
\exp\left\{\ii\oint\Lint_{\rm I}\di x\right\}\right>,
\end{equation}
%
where $\Tc$ denotes the special time-ordering operator, which orders the
operators according to a time parameter running along the time contour ${\cal
  C}$. The l.h.s. of eq. (\ref{cont.exp.}) is written in the Heisenberg
representation, whereas the r.h.s. is given in the interaction (${\rm I}$)
representation. Here and below, the subscript "I" indicates the {\em
  interaction} picture. Functions with $n$ points can be expressed in terms of
products of other multi-point functions contour integrated over internal
coordinates. Ultimately, one likes to express the multi-point functions of
interest in terms of one- and two-point functions (Wick's linked cluster
expansion), i.e.  in terms of classical fields, classical source-currents,
propagators and self-energies. Note that at this level the contour is not a
contour in the complex plane, as the figure may suggest, but rather it runs
along {\em real} time arguments. It is through the placement of external
points on the contour that the contour ordering obtains its particular sense.

In certain calculations, e.g., in those that apply the Fourier and Wigner
transformations, it is necessary to decompose the full contour into its two
branches---the {\em time-ordered} and {\em anti-time-ordered} branches. One
then has to distinguish between the physical space-time coordinates $x,\dots$
and the corresponding contour coordinates $x^{\cal C}$ which for a given $x$
take two values $x^-=(x^-_{\mu})$ and $x^+=(x^+_{\mu})$ ($\mu\in\{0,1,2,3\}$)
on the time ordered and anti-time ordered branches, respectively (see
figure 1).
Closed real-time contour integrations can then be decomposed as
%
\begin{eqnarray}
\oint\di x^{\cal C} \dots =\int_{t_0}^{\infty}\di x^-\dots
+\int^{t_0}_{\infty}\di x^+\dots
=\int_{t_0}^{\infty}\di x^-\dots -\int_{t_0}^{\infty}\di x^+\dots, 
\end{eqnarray}
%
where only the time limits are explicitly given.  Thus, the anti-time-ordered
branch acquires an extra minus sign if integrated over physical times. For any
two-point function $F$, the contour values are defined as
%
\begin{equation}\label{Fij}
F^{ij}(x,y):=F(x^i,y^j), \quad i,j\in\{-,+\},\,\,\,
\end{equation}
%
on the different branches of the contour. The contour $\delta$-function is
defined as
%
\begin{equation}\label{C-delta}
\dc (x^i,y^j) = \sigma^{ij} \delta^4(x-y),\quad
\sigma^{ij}=\left(\begin{array}{cc}1&0\\0&-1\end{array}\right)
\end{equation}
%
where, $\sigma_{ik}$ is the $\sigma_3$-Pauli matrix.  For any multi-point
function, the external point $x_{max}$, which has the largest physical time,
can be placed on either branch of the contour without changing the value,
since the contour-time evolution from $x_{max}^-$ to $x_{max}^+$ provides
unity. Therefore, one-point functions have the same value on both sides of the
contour.

Due to the change of operator ordering, genuine multi-point functions are
discontinuous in general, when two contour coordinates become identical. In
particular, two-point functions like $\ii F(x,y)=\left<\Tc {\widehat
A(x)}\medhat{B}(y)\right>$ become\footnote{Quite commonly, like in
refs. \cite{Kad62,Dan84}, the notation $ F=\left(\begin{array}{cc} F^c& F^<\\
F^>& F^a\end{array}\right)$ is used for two-point functions instead of
(\ref{Fxy}). We prefer the more flexible $\{-+\}$ labeling of contour points.}
%
\begin{eqnarray}\label{Fxy}
\hspace*{-1cm}\ii F(x,y) &=&
\left(\begin{array}{ccc} 
\ii F^{--}(x,y)&&\ii F^{-+}(x,y)\\[3mm]
\ii F^{+-}(x,y)&&\ii F^{++}(x,y)
\end{array}\right)=
\left(\begin{array}{ccc} 
\left<{\cal T}\medhat{A}(x)\medhat{B}(y)\right>&\hspace*{5mm}&
\left<\medhat{B}(y)\medhat{A}(x)\right>\\[5mm]
\left<\medhat{A}(x)\medhat{B}(y)\right>
&&\left<{\cal T}^{-1}\medhat{A}(x)\medhat{B}(y)\right>
\end{array}\right), 
\end{eqnarray}
%
where ${\cal T}$ and ${\cal T}^{-1}$ are the usual time and anti-time ordering
operators.  Since there are altogether only two possible orderings of the two
operators, in fact given by the Wightman functions $F^{-+}$ and $F^{+-}$,
which are both continuous, not all four components of $F$ are independent.
 From eq. (\ref{Fxy}) follow relations between non-equilibrium and usual
retarded and advanced functions
%
\begin{eqnarray}\label{Fretarded}
F^R(x,y)&=&F^{--}(x,y)-F^{-+}(x,y)=F^{+-}(x,y)-F^{++}(x,y)\nonumber\\
&:=&\Theta(x_0-y_0)\left(F^{+-}(x,y)-F^{-+}(x,y)\right),\nonumber\\
F^A (x,y)&=&F^{--}(x,y)-F^{+-}(x,y)=F^{-+}(x,y)-F^{++}(x,y)\nonumber\\
&:=&-\Theta(y_0-x_0)\left(F^{+-}(x,y)-F^{-+}(x,y)\right),
\end{eqnarray}
%
where $\Theta(x_0-y_0)$ is the step function of the time difference. 

Discontinuities of a two-point function may cause problems for
differentiations, in particular, since they often occur simultaneously in
products of two or more two-point functions. The proper procedure is, first,
with the help of eq.  (\ref{Fretarded}) to represent the discontinuous parts
in $F^{--}$ and $F^{++}$ by the continuous $F^{-+}$ and $F^{+-}$ times
$\Theta$-functions, then to combine all discontinuities, e.g. with respect to
$x_0-y_0$, into a single term proportional to $\Theta(x_0-y_0)$, and finally
to apply the differentiations.  One can easily check that in the
following particularly relevant cases
%
\begin{eqnarray}
\label{diffrules0}
&&
\oint\di z\left(
F(x^i,z)G(z,x^j) - G(x^i,z)F(z,x^j)\right), 
\\ 
\label{diffrules}
&&
\frac{\partial}{\partial x_{\mu}}
\oint\di z\left(
F(x^i,z)G(z,x^j)+G(x^i,z)F(z,x^j)\right),
\\ \label{diffrules1}
&&\left[\left(\frac{\partial}{\partial x_{\mu}}
              -\frac{\partial}{\partial y_{\mu}}\right)
\oint\di z \vphantom{\frac{\partial}{\partial x_{\mu}}
              -\frac{\partial}{\partial y_{\mu}}}
\left(F(x^i,z)G(z,y^j)-G(x^i,z)F(z,y^j)\right)\right]_{x=y}
\end{eqnarray}
%
{\em all discontinuities exactly cancel}. Thereby, these values are
independent of the placement of $x^i$ and $x^j$ on the contour, i.e. the
values are only a function of the physical coordinate $x$.

Boson fields may take non-vanishing expectation values of the field
operators $\Pba (x)=\left<\Pa \right>$, called {\em classical} fields. The
corresponding equations of motion are provided by the ensemble
averaging of the operator equations of motion (\ref{eqmotion})
%
\begin{eqnarray}
\label{eqmotion1}
S_{x}\Pba (x)= - J (x) ,\quad{\rm or}\quad
\Pba (x)=\Pba^0(x) -\oint\di y\Gr^{0}(x,y) J (y).
\end{eqnarray}
%
Here $J (x)=\left<\j (x)\right>$, while $\Pba^0 (x)=\left<\Pa_{\scr{I}}
  (x)\right>$ is the freely evolving classical field which starts from $\Pba
(t_0,{\vec x})$ at time $t_0$. Thereby, $\Ga^{0} (x,y)$ is the free contour
Green's function
%
\begin{eqnarray}\label{G0-int}
\ii\Ga^{0} (x,y)=
\left<\Tc\Ph_{\scr{I}} (x)\Ph_{\scr{I}}^{\dagger} (y)\right> 
-\Pba^0(x)(\Pba^0(y))^*.
\end{eqnarray}
%
The reader can easily verify that the contour form (\ref{eqmotion1}) is
equivalent to the standard retarded form of the classical field equations due
to eq. (\ref{Fretarded}) and the fact that $J (x)$ and $\Pba (x)$ are
one-point functions, which have identical values on both sides of the contour.
The free propagator is resolvent of equation
%
\begin{eqnarray}
\label{G0-eq.}
S_{x}\Gr^{0}(x,y)
&=&\dc(x,y)
\end{eqnarray}
%
on the contour, where $\dc (x^i,y^j)$ is the contour $\delta$-function
(\ref{C-delta}). Subtracting the classical fields via
%
\begin{eqnarray}
\label{phi-phi_cl}
\Ph =\Pba +\Pt , 
\end{eqnarray}
%
we define the full propagator in terms of quantum-fluctuating parts
$\Pt$ of the fields 
%
\begin{eqnarray}\label{Ga1}
\ii\Gr(x,y) =
\left<\Tc\Pt (x)\Ptd(y)\right>
=\left<\Tc\Ph (x)\Phd (y)\right> -\phi (x)\phi^{*} (y)
=\left<\Tc\Ph (x)\Phd (y)\right>_c .
\end{eqnarray}
%
Here and below, the sub-label $"c"$ indicates that uncorrelated parts
are subtracted. In terms 
of diagrams it implies, that the corresponding expectation values are
given by sums of entirely {\em connected} diagrams.

Averaging the operator equations of motion (\ref{eqmotion}) multiplied by
$\Pad(x)$ and subtracting classical-field parts, one obtains the equation of
motion for the propagators as
%
\begin{eqnarray}\label{Dyson0}
S_{x}\Gr (x,y) &=& \dc(x,y)+\ii\left<\Tc {\j}(x) \Ptd(y)\right>_c ,
\end{eqnarray}
%
where the contour $\delta$-function appears due to the contour ordering of the
operators in $\Gr(x,y)$.  

Eq. (\ref{Dyson0}) is still exact and accounts for the full set of initial
correlations contained in $\medhat{\rho}_0$. In order to proceed, one may
suggest that the typical interaction time $\tau_{\scr{int}}$ for the change of
the correlation functions is significantly shorter than the typical relaxation
time $\tau_{\scr{rel}}$, which determines the system evolution, when one
neglects those initial correlations. Then, describing the system at times
$t-t_0 \gg \tau_{\scr{int}}$, one can neglect the initial correlations which
are supposed to be dying at the time $\sim \tau_{\scr{int}}$ in accordance
with the Bogolyubov's principle of the weakening of initial
correlations.\footnote{Actually, considering a dilute gas limit, Bogolyubov
suggested the weakening of all the correlations, whereas we use a weaker
assumption on the weakening of only short-range ($\sim \tau_{\scr{int}}$)
correlations, cf. \cite{Bogolyubov}.}  As a result one can apply the standard
Wick decomposition dropping higher order correlations for the driving terms on
the r.h.s. of both equations of motion (\ref{eqmotion1}) and
(\ref{Dyson0}). Then both driving terms can be expressed solely as functionals
of the classical fields and one-particle propagators rather than on higher
order correlations. In particular the driving term of (\ref{Dyson0}) can then
be expressed through the proper self-energy $\Sa$
%
\begin{eqnarray}
\label{eqmotion2}
\ii\left<\Tc {\j}(x) \Ptd(y)\right>_c
&=&\ii\left<\Tc \left[
\exp\left\{\ii \oint \di z \Lint_{\scr{I}}\right\}
\j_{\scr{I}}(x)\Ptd_{\scr{I}}(y)
\right]\right>_{c}
\nonumber
\\
&=&\ii\oint\di z\left<\Tc 
\frac{\partial }{\partial \Pt_{\scr{I}}(z)} \left[
\exp\left\{\ii \oint \di z' \Lint_{\scr{I}}\right\}
\j_{\scr{I}}(x)\right]\right>_{c1}
\left<\Tc\Pt(z) \Ptd(y)\right>\nonumber\\ 
&=&\oint\di z\Sa (x,z)\Ga (z,y). 
\end{eqnarray}
%
The second line of eq. (\ref{eqmotion2}) results from the fact that the
expectation value is connected and, therefore, in the first line,
$\Ptd_{\scr{I}} (y)$ has to be contracted with one of the operators $\PtI(z)$
occurring in $\exp\left\{\ii \oint \di z'
\Lint_{\scr{I}}\right\}{\j_{\scr{I}}}(x)$. The expression for the self-energy
$\Sa$ deserves attention, since we have separated the {\em full} propagator in
(\ref{eqmotion2}). In order to achieve the proper counting of terms, $\Sa$ has
to be {\em one-particle irreducible} (label $c1$), i.e. the corresponding
diagram cannot be split into two pieces, which separate $x$ from $z$ by
cutting a single propagator line.  Obviously, $\Se$ may have singular
($\delta$-functional) one-point parts and genuine two-point parts (the latter
are given by all connected {\em one-particle irreducible diagrams} of the
current--current correlator)
%
\begin{eqnarray}
\label{Se-definition}
-\ii\Sa (x,y)=
\left<\Tc\frac{\partial^2\ii\Lint(x)}{\partial\Ph \partial\Phd}\right>_{c}
\dc(x,y)
-\left<\Tc\j (x)\j^{\dagger}(y)\right>_{1c}, 
\end{eqnarray}
%
in the Heisenberg picture. 

With the help of eqs (\ref{Dyson0}) and (\ref{eqmotion2}), one
recovers the Dyson equation in the differential form 
%
\begin{eqnarray}\label{Dyson}
S_{x}\Gr (x,y) &=& \dc(x,y)+\oint \di z \Se (x,z)\Gr (z,y),
\\
\label{DysonI}
\left(S_y\right)^*\Gr(x,y)&=&\dc(x,y)+\oint \di z \Gr(x,z)\Se(z,y). 
\end{eqnarray}
%
Using the resolvent property of the free propagator, we can write
down these equations also in the integral form.

In diagrams, free and full classical fields are represented by ''pins'' with
cross and ''o-cross'' as heads
%
\begin{eqnarray}
\phi^0(x)
=\GPb\hspace*{1cm}\phi(x)=\GPbfull,\hspace*{1cm}
\vphantom{\int}
\end{eqnarray}
while free and full propagators are given by thin and thick lines,
respectively, 
\begin{eqnarray}
\ii\Gr^0(x,y)=\GGxy\quad\hspace*{1cm}\ii\Gr(x,y)=\GGfullxy
\vphantom{\int}
\end{eqnarray}
%
Thereby, complex fields carry a sense, the arrow always pointing towards the
$\Ph$ in the contour ordered expressions. In diagrammatic representation, the
classical-field equations (\ref{eqmotion1}) and Dyson equations (\ref{Dyson})
are given by
%
\begin{eqnarray}
\GPbfull&=&\GPb + \GGJ\vphantom{\displaystyle\int_A^B},\\
\GGfull&=&\GG + \Dysonself\vphantom{\displaystyle\int_A^B}
\end{eqnarray}
%
with the one- and two-point functions $\ii J(x)$ and $-\ii\Se(x,y)$ 
as driving terms.

The averaged values of conserved quantities can be expressed in terms of the
one- and two-point functions introduced so far. Averaging the operator value of
the energy--momentum tensor of eq.  (\ref{E-M-pi-tensor}), we arrive at
%
\begin{eqnarray}
\label{E-M-G-tensor}
\nonumber
\Theta^{\mu\nu} (x)=:
\left<\widehat{\Theta}^{\mu\nu}_{\scr{(alt.)}}\right> &=&
\frac{1}{2}\suma
\left[
\left(\left(\widehat{p}^{\nu}_x\right)^* 
+ \widehat{p}^{\nu}_y\right)
\left(\left(\widehat{p}^{\mu}_x\right)^* 
+ \widehat{p}^{\mu}_y\right) 
\left(\vphantom{\frac{1}{2}}
\Pba^*(x) \Pba(y) +
\ii\Ga^{\scr{sym}} (y,x)\vphantom{\frac{1}{2}}\right)
\right]_{x=y} 
\nonumber
\\
&+&
g^{\mu\nu}
\left({\cal E}^{\scr{int}}(x)-
{\cal E}^{\scr{pot}}(x)
\right),
\end{eqnarray}
%
where ${\cal E}^{\scr{int}}(x)=-\left<\Lint(x)\right>$, and 
the potential energy density becomes
%
\begin{eqnarray}
\label{eps-pot1}
{\cal E}^{\scr{pot}}(x)
&=&\left<{\widehat{\cal E}}^{\scr{pot}}(x)\right>
=\displaystyle\frac{1}{2}\suma
\left\{
\vphantom{\oint}
- 
\left[\jba^*(x)\Pba(x) + \jba(x)\Pba^*(x)\right] 
\right. \nonumber
\\
&+&\displaystyle\left. \ii 
\oint\di
z\left[\Sa(x,z)\Ga(z,x)+\Ga(x,z)\Sa(z,x)\right]
\right\}. 
\end{eqnarray}
%
due to eqs (\ref{eps-pot}) and (\ref{eqmotion2}). Note that we do not
prescribe any contour indices to $x$ in the integral term of ${\cal
E}^{\scr{pot}}$, since actually this term is independent of the
contour placement of $x$ due to discontinuity property (\ref{diffrules}). 
The Noether current (\ref{c-pi-current}) takes the form
%
\begin{eqnarray}
\label{c-G-current} 
j^{\mu} (x) =:
\left<\medhat{j}^{\mu}_{\scr{(Noether)}}\right>
&=& e 
\left[
\left(\left(\widehat{p}^{\mu}_x\right)^* 
+ \widehat{p}^{\mu}_y\right) 
\left(\vphantom{\frac{1}{2}}
\Pba^*(x) \Pba(y) +
\ii\Ga^{\scr{sym}} (y,x)\right)
\right]_{x=y} .
\end{eqnarray}
%
In order to keep the expressions $\Theta^{\mu\nu}$ and $j^{\mu}$ charge
symmetric we have introduced the symmetric quantities
%
\begin{equation}
\label{F-sym} 
F^{\scr{sym}}(x,y)= 
\frac{1}{2}\left(F(x^-,y^+)+F(x^+,y^-)\right), 
\end{equation}
%
where $F(x,y)$ is any two-point function on the real-time contour. This is not
automatically provided by the variational methods leading to the Noether
energy--momentum tensor and current, as they only provide the vanishing of the
corresponding divergence. The integrated form of the conserved quantities has
to be adjusted such that charge symmetry is maintained, thus describing
contributions of both, particles and anti-particles, on equal footing. This
way one properly accounts for the modification of the vacuum polarization in
the medium, since the vacuum-polarization energy coincides with the zero point
energies of the field oscillations. The corresponding divergence has still to
be appropriately renormalized.

\section{Functionals $W$, $\Gamma$ and $\Phi$}\label{sect-W-Phi}

The standard generating functional for connected $n$-point functions is given
by the logarithm of the expectation value of the time-evolution operator with
external one-point sources $\eta(x)$ and $\eta^*(x)$
%
\begin{eqnarray}
\label{W1}
&&W\left\{\eta,\eta^*\right\}
=-\ii\ln\left<
{\cal T} \exp\left[ \ii\int\di x
\left[\eta(x)\Phd(x)+\eta^* (x)\Ph(x)\right] 
\right]\right>\nonumber\\
&&=-\ii\ln\left<
\exp\left[-\ii\int\di t \Hh^0_{\scr{I}}\right]
{\cal T} \exp\left[ \ii\int\di x
\left( \Lint_{\scr{I}}
+
\left[\ea(x)\Phd_{\scr{I}}(x)+\ea^* (x)\Ph_{\scr{I}}(x)\right] 
\right)\right]\right>_{\scr{I}},\nonumber
\end{eqnarray}
%
where $\Hh^0_{\scr{I}}$ is the free Hamiltonian of the system in the
interaction representation.  Since the functional dependence concerns only the
external sources $\eta(x)$, the operator part can be cast into different
pictures, such as the Heisenberg or interaction ones. The latter establishes
the perturbation expansion.  Here and below, $\left<\dots\right>$ denotes a
trace over all states, which includes the ensemble average over the density
operator $\medhat{\rho}_0 = \medhat{\rho}(t_0)$ at initial time $t_0$ (cf. eq.
(\ref{corrfct})).

It is advantageous to introduce a scale function $\lambda(x)$, which scales
interaction vertices of the interaction Lagrangian density $\Lint$ at
space-time coordinate $x$ and defines a $\lambda(x)$ dependent interaction
Lagrangian density
%
\begin{equation}
\label{L_l}
\Lint_{\lambda}=
\lambda(x)\Lint
\left\{ \Phd (x),\Ph (x)\right\} .
\end{equation}
%
This scaling provides the clue to the proof of the diagrammatic
representation of the auxiliary functional $\Phi$ in terms of closed diagrams.
 
We generalize the above $W$ functional to the real-time contour ${\cal C}$.
Following Luttinger and Ward \cite{Luttinger} and Baym \cite{Baym}, we
also extend it to include external bilinear sources $\Ka(x,y)$, besides the
interaction scale $\lambda(x)$, 
%
\begin{eqnarray}
\label{W}
\!\!\!\!\!\!
&&W\left\{\eta,K,\lambda \right\}
=-\ii\ln\left<
\exp\left[-\ii\oint\di x \Hh_{\scr{I}}^{0} \right] 
\Tc \exp\left\{ \ii\oint\di x
\left( \vphantom{\suma}
\lambda\Lint_{\scr{I}}
\right.\right.\right.
\nonumber\\
&&+
\left.\left.\left.
\suma\left[
\ea(x)\Ph_{\scr{I}}^{\dagger}(x)+\ea^* (x)\Ph_{\scr{I}}(x)
+\ii\oint\di y \Ph_{\scr{I}}(x)\Ka(x,y)
\Ph_{\scr{I}}^{\dagger}(y)
\right]\right)
\right\}\right> .
\end{eqnarray}
%

While $n$-th order functional variations of (\ref{W}) with respect to
$\ea$ generate the {\em connected} $n$-point functions, the first
functional variation with respect to $K(x,y)$ gives the total two-point
propagator including disconnected pieces
%
\begin{eqnarray}
\label{varW}
\delta W\left\{\eta,K,\lambda \right\}&=&
\suma \left[ 
\oint\di x\left[\Pba^*(x)\delta \ea(x)+\Pba(x)\delta \ea^*(x)
\right]
\right. 
\nonumber
\\
&&\left.  
+\oint\di x\oint\di y\left[\Pba(x)\Pba^*(y)
+\Ga(x,y)\right]\delta \Ka(y,x)
\right]
\nonumber
\\
&-&\oint\di x {\cal E}^{\scr{int}}_\lambda(x)\delta\lambda(x)/\lambda(x). 
\end{eqnarray}
%
Here, $\Pba$ and $\Ga$ denote the classical fields and propagators,
respectively, which now implicitly depend on $\lambda$.  The term, resulting
from the variation of the interaction scale function $\lambda(x)$, defines a
one-point function ${\cal
E}^{\scr{int}}_\lambda(x)=-\left<\Lint_{\lambda}(x)\right>$, which agrees with
the corresponding expectation value of (\ref{eps-pot}) but for the scaled
Lagrangian.

The step towards a functional that depends on the classical fields $\Pba$ and
propagators $\Ga$ rather than on the external sources $\ea$ and $\Ka$ is
provided by the double Legendre transformation of $W$ to
$\Gamma\{\phi,\phi^*,\Gr,\lambda \}$ given as
%
\begin{eqnarray}
\label{gamma-pert}
\Gamma\{\phi,\phi^*,\Gr,\lambda \}
&=&
W\{\eta,K,\lambda \}
-  \suma\left[
\oint\di x
\left[\ea^*(x)\Pba(x)+\ea(x)\Pba^*(x)\right]
\right. 
\nonumber
\\
&&\left.
+\oint\di x\di y\left[\Pba(y)\Pba^*(x)+\Ga(y,x)\right]\Ka(y,x)
\right]. 
\end{eqnarray}
%
Here, the sources $\ea$ and $\Ka$ have to be expressed through
$\Pba$ and $\Ga$. Apart from the $\delta \lambda$
dependences, the functional variation
%
\begin{eqnarray}
\label{varPhi}
\delta \Gamma\left\{\phi,\phi^*,\Gr,\lambda \right\}&=&
-  \suma\left[
\oint\di x\left[\ea^*(x)\delta \Pba(x)+\ea(x)\delta \Pba^*(x)
\right]
\right.
\nonumber\\
&+&
\left.
\oint\di x\oint\di y\left[\delta\Pba(y)\Pba^*(x)+\Pba(y)\delta\Pba^*(x)
+\delta \Ga(y,x)\right]\Ka(y,x)
\right] 
\nonumber\\
&-&
\oint\di x {\cal E}^{\scr{int}}_{\lambda}(x)\delta\lambda(x)/\lambda(x)
\end{eqnarray}
%
vanishes at vanishing external sources $\ea$ and $\Ka$. The latter together
with the condition $\lambda=1$ corresponds to the physical solution. Note also
that for the physical solution, i.e. at $\ea = \Ka = 0$, the values of the two
functionals $\Gamma$ and $W$ are identical. Indeed, the variations
%
\begin{eqnarray}
\label{varG/phi}
\delta \Gamma / \delta \Pba = 0, \,\,\,
\delta \Gamma / \delta \Pba^* = 0, \,\,\,
\delta \Gamma / \delta \Ga = 0 
\end{eqnarray}
%
provide us with equations of motion for classical fields (\ref{eqmotion1}) and
the complex conjugated one, as well as Dyson's equation (\ref{Dyson}).

Following Luttinger, Ward \cite{Luttinger}, and later by Cornwell et al.
\cite{Cornwell}, who  used path-integral methods for case of the imaginary-time
formulation of equilibrium systems, we represent our $\Gamma$ functional
related to the real-time quantities in the form
%
\begin{eqnarray}\label{Gammafull}
\displaystyle &&\Gamma\{\phi,\phi^*,\Gr,\lambda \}
=\displaystyle \Gamma^0+\displaystyle
 \oint \di x \Lg^0\{\phi,\partial_\mu\phi\}
\nonumber 
\\
&&+ \ii \suma \left[
\ln\left(1-\odot\Ga^{0}\odot\Sa\right)
+\odot\Ga\odot\Sa \right]
+\displaystyle \Phi\left\{\phi,\phi^*,\Gr,\lambda \right\}, 
\label{PHIfunct}
\end{eqnarray}
%
this way defining the auxiliary functional $\Phi\left\{\phi,\phi^*,\Gr,\lambda
\right\}$\footnote{Please note that our $\Phi$ is different from the auxiliary
quantity $\Gamma_2$ defined by Cornwell et. al. \cite{Cornwell} to the extent
that their $\Gamma_2$ is void of zero- and one-loop terms. These terms are
rather included in defining a full classical Lagrangian that depends on the
classical fields and tad-poles, while we have placed all interaction parts
into $\Phi$.}. Constructing $\Phi$ solely in terms of the fields and
one-particle propagators complies with the assumption of ignoring higher order
correlations.  The $\Gamma^0$ and $\Lg^0$ parts, where $\Lg^0$ is
the free {\em classical} Lagrangian function, represent the non-interacting
parts of $\Gamma$. Thereby, the $\Gamma^0$ term solely depends on the
unperturbed propagator $\Gr^0$ and hence is treated as a constant with respect
to functional variations of $\Gamma$.  The $\ln(\dots)$ is understood in the
functional sense, i.e. by a series of $n$-folded contour convolutions, denoted
by the $\odot$-symbol, formally resulting from the Taylor expansion of the
$\ln(1+x)$ at $x=0$. This $\ln$-term accounts for the change of $\Gamma$
due to the self-energies of the particles. The $\Gamma^0$, $\Lg^0$ and $\ln$
terms in eq. (\ref{PHIfunct}) account for the one-body components in the
$\Gamma$. The remaining $\Ga\odot\Sa$ and $\Phi$ terms correct for the true
interaction energy part of $\Gamma$. As shown in the next section, the form
(\ref{Gammafull}) presents a re-summation of the corresponding perturbative
expansion of the value of $\Gamma$ (cf. eq. (\ref{gamma-pert})) in terms of
full classical fields and propagators.

The specific form (\ref{Gammafull}) has important functional properties, which
provide us with a number of useful relations. Functional variation of
$\Gamma\left\{\phi,\phi^*,\Gr,\lambda\right\}$ in the form of eq.
(\ref{PHIfunct}) leads to
%
\begin{eqnarray}
\label{vargamma}\hspace*{-1cm}
\displaystyle \delta\Gamma \left\{\phi,\phi^*,\Gr,\lambda\right\}
&=&\displaystyle
\suma\left\{
\oint\di x 
\left[\delta\Pba (x)(\So_x )^{\ast} \Pba^*(x)+\delta\Pba^*(x)\So_x
\Pba(x)\right]\right.
\nonumber 
\\
\hspace*{-1cm}&&\hspace*{2cm}
- 
\ii  \left(
\odot \frac{1}{1-\odot\Ga^{0}\odot\Sa}\odot\Ga^{0} - \odot\Ga
\right)\odot\delta\Sa
\nonumber 
\\
\hspace*{-1cm}&&\hspace*{2cm}
+\displaystyle\left.
 \ii  \oint \di x \di y \Sa(x,y) \delta\Ga(y,x) \right\}
+\delta\Phi\left\{\phi,\phi^*,\Gr,\lambda\right\}.
\end{eqnarray}
%
Here $\delta\Sa$ is understood as a variation induced by $\delta\Ga$,
$\delta\Pba$, $\delta\Pba^*$, and $\delta\lambda$, respectively. Eqs
(\ref{varPhi}) and (\ref{varG/phi}) imply the following variational rules for
the auxiliary $\Phi$ functional
%
\begin{eqnarray}
\label{varphi'}
&\delta&\Phi\left\{\phi,\phi^*,\Gr,\lambda\right\}=\displaystyle
\suma\left\{
\oint\di x 
\left[\jba^*(x)\delta\Pba(x)+\jba(x)\delta\Pba^*(x)\right]
\nonumber\right. \\
&-&\displaystyle\left.
 \ii  \oint \di x \di y
\Sa(x,y)\delta\Ga(y,x)\right\} 
-\displaystyle
\oint\di x{\cal E}^{\scr{int}}(x)\delta\lambda(x),
\end{eqnarray}
%
or 
%
\begin{eqnarray}\label{varphdl}
\ii\jba(x)&=&\frac{\delta\ii \Phi}{\delta \Pba^{\ast} (x)},\\
\label{varphdl1}
-\ii \Sa(x,y)&=&\frac{\delta\ii \Phi}{\delta \ii\Ga(y,x)}\times
\left\{
\begin{array}{ll}
2\quad&\mbox{for neutral bosons,}\\[1.5mm]
1\quad&\mbox{for charged bosons,}
\end{array}\right.\\
\label{varphdl2}
- {\cal E}^{\scr{int}}(x)&=& 
\frac{\delta\ii \Phi}{\delta \ii\lambda (x)}.
\end{eqnarray}
%
The virtue of the functional form (\ref{PHIfunct}) is that these requirements
can be met simultaneously and that there exists a unique form of $\Phi$, for
which the three derived quantities---the one-body source current $\jba(x)$,
the two-body self-energy $-\ii \Sa(x,y)$ and the interaction energy density
${\cal E}^{\scr{int}}(x)$---take their physical values at the physical
solutions of the equations of motion (\ref{Dyson}) and (\ref{eqmotion1}). This
will become clear in more detail in the next section, where we discuss the
diagrams defining the various functionals. In its turn, $\Phi$ can be seen as a
{\em generating} functional for the source terms $\jba$ of classical fields
and self-energies $\Sa$ for the set of Dyson equations. Therefore,
approximation schemes can be defined through a particular approximation to
$\Phi$. Thereby, the invariance properties of $\Phi$ play a central role to
define conservation laws for the approximate dynamics.

It is important to emphasize that we do all functional variations
independently of any place on the contour. Thus, different contour times are
considered as independent, even though they may refer to the same physical
time\footnote{ The fact that for the physical solutions the components of
$\Ga$ on the different branches of the contour are not independent
(cf. (\ref{Fretarded})), has no importance for the variational procedure. The
reason is that rules (\ref{Fretarded}) only apply to the physical $\Ga$ and
$\Pba$, which are provided by the stationary ``points'' of the variational
principle, i.e.  solving the equations of motion (\ref{eqmotion1}),
(\ref{Dyson}) and (\ref{DysonI}).}. In principle, all variational
considerations given in this section apply to any kind of time contour,
even to
non closed and complex ones as well as to any operator ${\medhat \rho}_0$
defining the averaging $\left<\dots\right>=\Tr\left\{\dots{\medhat
\rho}_0\right\}$ including the unit operator, as used in Matsubara's
imaginary-time formalism. For a particular choice of ${\medhat \rho}_0$ and
of a contour the {\em physical} values of $W$ and $\Gamma$ are identical for
the corresponding physical solutions along this contour. In the imaginary-time
method the value $\Gamma=W$ takes the meaning of the thermodynamic partition
sum. In this paper we concentrate on the non-equilibrium closed real-time
formalism for which $\Tr{\medhat \rho}_0=1$, and therefore the {\em physical}
values of $W$ and $\Gamma$ trivially vanish, i.e.
$$W=\Gamma=\Phi=0 \quad {\mbox{\rm for physical solutions of 
$\Ga$, $\Pba$, $\Pba^*$ }}\quad {\mbox{on the contour}}.$$

An important comment must be given at this stage. One should clearly
distinguish between the {\em functional} form of a functional, which acquires
its meaning through variational methods, and the {\em physical} value that
functional takes once the physical solutions of the equations of motion are
inserted. For instance, two functionals $W$ and $\Gamma$ are completely
different in their functional meaning, while they take the same physical value
for the physical solution.  Therefore, for all functionals the functional
dependences are explicitly given in braces.  Our strategy below will be first
to perform general variations of $\Gamma$ and $\Phi$, allowing non-physical
values of $\Ga$, $\Pba$, $\Pba^*$ and $\lambda$, and only then to put them to
their physical values. This way, a number of important relations between
Green's functions, self-energies and mean fields will be obtained.

\section{Diagrams for $\Gamma$, $\Phi$ and 
${\cal E}^{\scr{int}}_{\lambda}(x)$}\label{Diagrams}

According to eqs (\ref{varPhi}) and (\ref{varphdl2}), we have 
%
\begin{eqnarray}\label{ep-phi}
-\oint\di x {\cal E}^{\scr{int}}(x)=
\left[\lambda\frac{\di}{\di \lambda}
\Gamma\{\phi\{\lambda\},\phi^*\{\lambda\},\Gr\{\lambda\},\lambda\}
     \right]_{\lambda=1}
=\left[\lambda\frac{\partial}{\partial\lambda}
\Phi\{\phi,\phi^*,\Gr,\lambda\}\right]_{\lambda=1},
\end{eqnarray}
%
where now $\lambda$ is treated as a global scale parameter (note
that only a partial derivative is applied to $\Phi$, i.e. the $\Pba$,
$\Pba^*$ and $\Ga$ values are kept constants.). In the perturbation
theory, the diagrammatic rules to calculate the one-point 
function ${\cal E}^{\scr{int}}(x)$ are straightforward
\unitlength=.8cm
%
\begin{eqnarray}\label{Eint-diag}
-\ii{\cal E}^{\scr{int}}(x)
=\ii\left<\Tc\LintI(x)\exp\left[\ii\oint\di x'\LintI(x')\right]\right>
=\sum_{n_\lambda}\Dclosedone{c}{\thinlines}.\vhight{.8}
\end{eqnarray}
%
Here the diagram symbolically denotes all connected (label $c$) closed
perturbation-theory diagrams generated by expanding the exponential function
in (\ref{Eint-diag}).  The full dot denotes the external point $x$ which is
not integrated out. Integrating (\ref{ep-phi}) with respect to
$\lambda$, we define the quantity $\ii{\bar \Gamma}$ given by 
the following perturbative diagrammatic representation  \unitlength=.8cm
%
\begin{eqnarray}\label{Gamma-pert-diag}
\ii{\bar\Gamma}\{\phi^0,\phi^{0*},\Gr^0,\lambda \}
= \ii \Gamma^0\left\{\Gr^0\right\}
+ \ii \oint \di x \Lg^0\{\phi^0,\partial_\mu\phi^0\}
+\sum_{n_\lambda}\frac{1}{n_\lambda}
\Dclosed{c}{\thinlines},\vhight{.8}
\end{eqnarray}
%
where the integration constants have been chosen such that for physical
solutions ${\bar\Gamma}=\Gamma$.  One can see that each diagram contributing
to $\bar\Gamma$ has to be weighted with its inverse number of vertices
$1/n_\lambda$, due to the formal $\lambda$-integration of (\ref{Eint-diag}).
It is important to realize that due to these global factors such a set of
diagrams {\em is not resumable} in the standard diagrammatic
sense\footnote{Diagrammatic re-summation implies that sub-diagrams with the
  same external structure (i.e. same number of external points and types of
  propagators to be attached at each external point) can be summed up to give
  a total re-summed expression that can then be embedded into more complicated
  diagrams, e.g. self-energy insertions can be re-summed to full Green's
  functions.}.  Also $\bar\Gamma$ in the form of eq.
(\ref{Gamma-pert-diag}) is a functional of $\phi^0,\phi^{0*},\Gr^0,\lambda$
rather than of $\phi,\phi^*,\Gr,\lambda$, as needed for $\Gamma$ and $\Phi$ as
discussed in the previous section. However, we can arrive at the required
functional dependence of $\Gamma$ as follows.  The expression
(\ref{Eint-diag}) for $-\ii{\cal E}^{\scr{int}}(x)$ can be re-summed and
entirely expressed in terms of full classical fields and full propagators.
The re-summed diagrams are then void of any self-energy insertions and
therefore have to be {\em two-particle irreducible}
%
\begin{eqnarray}\label{Eint-diag1}
-\ii{\cal E}^{\scr{int}}(x)
=\sum_{n_\lambda}\Dclosedone{c2}{\thicklines}.\vhight{.8}
\end{eqnarray}
%
Diagrams of class $c2$ cannot be decomposed into two pieces by
cutting two propagator lines. The formal integration of the last equality in
(\ref{ep-phi}) with respect to $\lambda$ keeping $\phi$ and $\Gr$ constant
provides the diagrammatic expression for $\Phi$ in terms of full Green's
functions and classical fields. Therefore,
$\ii\Gamma\left\{\phi,\phi^*,\Gr,\lambda\right\}$ can be expressed in terms of
the following diagrams (cf. eq.  (\ref{PHIfunct}))
%
\begin{eqnarray}\label{keediag}
&&\ii\Gamma\left\{\phi,\phi^*,\Gr,\lambda\right\} = \ii
\Gamma^0\left\{\Gr^0\right\}  
+ \ii \oint \di x \Lg^0\{\phi,\partial_\mu\phi\} 
\nonumber 
\\
&&\hspace*{5mm}
+\suma
\left\{\vhight{1.6}\right.
\underbrace{\sum_{n_\Se}\vhight{1.6}\frac{1}{n_\Se}\GlnG0Sa}
_{\displaystyle -\ln\left(1-\odot\Ga^{0}\odot\Sa\right)}
\underbrace{-\vhight{1.6}\GGaSa}
_{\displaystyle -\odot\Ga\odot\Sa\vphantom{\left(\Ga^{0}\right)}}
\left.\vhight{1.6}\right\}
\underbrace{+\vhight{1.6}\sum_{n_\lambda}\frac{1}{n_\lambda}
\Dclosed{c2}{\thicklines}}
_{\displaystyle\vphantom{\left(\Ga^{0}\right)} +\ii\Phi}.
\end{eqnarray}
%
Here $n_\Se$ counts the number of $\Sa$ insertions in the ring diagrams
providing the $\ln$-terms, while for the closed diagrams of $\Phi$ the value
$n_\lambda$ counts the number of vertices building up the functional $\Phi$.
Contrary to the perturbative diagrams of $\ii{\bar\Gamma}$, cf. eq.
(\ref{Gamma-pert-diag}), here the diagrams contributing to $\Phi$ are given in
terms of full propagators $\Ga$ and full time-dependent classical fields
$\Pba$.  As a consequence, these diagrams have to be {\em two-particle
  irreducible} (label $c2$).  The latter property is required because of the
re-summations of ${\cal E}^{\scr{int}}(x)$.  This also matches the diagrammatic
rules for the re-summed self-energy $\Sa(x,y)$, which results from functional
variation of $\Phi$ with respect to any propagator $\Ga(y,x)$. In graphical
terms, this variation is realized by opening a propagator line in all diagrams
of $\Phi$.  The resulting set of thus opened diagrams must then be that of
proper skeleton diagrams of $\Sa$ in terms of {\em full propagators}, i.e.
void of any self-energy insertion.

The diagrammatic rules for $\Phi$, ${\cal E}^{\scr{int}}(x)$, $\jba$
and $\Sa$ are 
determined by the following steps:
\begin{itemize}
\item[(a)] For all bosonic fields in $\ii\Lint$, replace $\Pa$ by
$\Pba+\Pta$ in order to account for the classical fields
(cf. (\ref{phi-phi_cl})); 
\item[(b)] consider all possible pair contractions of the field operator
$\Pta(x)$ with $\Ptad(y)$ in the formal expressions
(\ref{Phi-diagrm})--(\ref{Pi-diagrm}) given below and
replace them by $\ii\Ga(x,y)$; 
\item[(c)] keep only those terms that correspond to two-particle
irreducible diagrams for $\Phi$ , i.e. which cannot be split into two
pieces by cutting two different propagator lines.
\end{itemize}
Further details are given in Appendix \ref{Diag-rules}. The diagrams of
$\ii\Phi$, $-\ii{\cal E}^{\scr{int}}(x)$ , $\ii\jba(x)$ and $-\ii\Sa(x,y)$
are then generated by applying the above general rules to the following formal
expressions
%
\begin{eqnarray}
\label{Phi-diagrm}
\hspace*{-1cm} 
\ii\Phi&=&\left<\Tc \exp\left(\ii\oint\di x' \Lint(x')\right) 
\right>_{2c\left\{\Ga\right\}}\nonumber\\[-2mm]&&\\[-2mm]\nonumber
&=&\sum_n \frac{1}{n!}\oint \di x_1\dots\di x_n
\left< \Tc 
\ii \Lint(x_1) \dots \ii \Lint(x_n)
\right>_{2c\left\{\Ga\right\}}, 
\\
-\ii{\cal E}^{\scr{int}}(x)&=&
\left<\Tc \ii \Lint(x)\exp\left(\ii\oint\di x' \Lint(x')\right) 
\right>_{2c\left\{\Ga\right\}} , 
\end{eqnarray}
%
%
\begin{eqnarray}
\label{J-diagrm}
\hspace*{-1cm} \ii\jba(x)&&
=\left<\Tc\frac{\delta}{\delta\Pba^*(x)} 
\exp\left(\ii\oint\di x' \Lint(x')\right) 
\right>_{2c\left\{\Ga\right\}},
\end{eqnarray}
%
%
\begin{eqnarray}
\label{Pi-diagrm}
\hspace*{-1cm} -\ii\Sa(x,y)
=\left<\Tc\frac{\delta^2}{\delta\Pta^\dagger(x)\delta\Pta(y)} 
\exp\left(\ii\oint\di x' \Lint(x')\right) 
\right>_{2c\left\{\Ga\right\}}, 
\end{eqnarray}
%
where the sub-label ${2c\left\{\Gr\right\}}$ refers to the above point
(c).

As an example, we quote the diagrams in neutral scalar $g \Ph^4/4!$ theory. Up
to two vertices, the functional $\Phi$ is given by the following
expressions
%
\begin{eqnarray}
\label{Phi/phi4}
\ii \Phi&=&\frac{-\ii g}{4!} \oint\di x
\left(\phi^4(x)+6\phi^2(x)
\left<\Pta(x)\Pta(x)\right>_c+
3\left<\Pta(x)\Pta(x)\right>_c^2\right)
\nonumber
\\[-2mm]&&\\[-2mm]
&+& 
\frac {1}{2} \left(\frac{-\ii g}{4!}\right)^2
\oint\di x\oint\di y \left(4\cdot4!\phi(x)\phi(y)
\left<\Pta(x)\Pta(y)\right>_c^3
+{4!}\left<\Pta(x)\Pta(y)\right>_c^4\right)+\dots,
\nonumber
\end{eqnarray}
%
or in terms of diagrams (cf. also Appendix \ref{Diag-rules}) 
%
\unitlength=.600mm
\begin{eqnarray}\label{phi4diagrams}\linethickness{3.pt}
\begin{array}{lccccccccccc}
\ii\Phi &=&  \gphifour 
&+&  \gloopphitwo 
&+& \geight
&+\displaystyle\frac{1}{2}\left\{\vphantom{\rule{0mm}{1.cm}}\right.&  
	\gphisand3phi 
&+&\gsand4
&\left.\vphantom{\rule{0mm}{1.cm}}\right\} 
	+\displaystyle\frac{1}{3}\dots\vphantom{\rule{0mm}{1.cm}}
\\[7mm]
&&\left[\frac{1}{4!}\right]
&&\left[\frac{1}{2\cdot2!}\right]
&&\left[\frac{1}{2^2\cdot 2!}\right]
&&\left[\frac{1}{3!}\right]
&&\left[\frac{1}{4!}\right]&
\end{array}\nonumber\\[-0.6cm]
\end{eqnarray}
%
The $1/n_\lambda$ factors are explicitly given, while the combinatorial factors
according to rule (vii) in Appendix A are given in square brackets below each
diagram. Functional derivatives with respect to $\phi$ (pins) and propagators
(full lines), cf. eq.(\ref{varphi'}), determine the source $J(x)$ of the
classical field and the self-energy $\Se(x,y)$, respectively,
%
\begin{eqnarray}
\label{J/phi4}
\begin{array}{rccccccccl}
\ii J(x)
&=&\gphithree &+& \gloopphione &+&
\gloopphih\;\;&+&\dots\vphantom{\rule{0mm}{1.cm}} ,\\[7mm]
&&\left[\frac{1}{3!}\right]
&&\left[\frac{1}{2}\right]
&&\left[\frac{1}{3!}\right]
&\\[6mm]
- \ii \Se(x,y)
&=& \sphitwo &+&
 \sloop &+&  
\sphisand2phi
&+&\ssand3&+\dots\vphantom{\rule{0mm}{1.cm}}\\[7mm]
&&\left[\frac{1}{2!}\right]
&&\left[\frac{1}{2}\right]
&&\left[\frac{1}{2!}\right]
&&\left[\frac{1}{3!}\right]
&
\end{array}
\end{eqnarray}
%
Small full dots define vertices which are to be integrated over, while
big full dots specify the external points $x$ or $y$; the first two
diagrams of $\Se(x,y)$ give the singular $\dc(x,y)$ parts arising from
classical fields and tad-poles.

\section{$\Phi$-Derivable Approximations and Invariances of $\Phi$ }

The expressions for $W$, $\Gamma$ and $\Phi$ given so far are exact and
represent a convenient formulation of the theory in terms of full propagators
and self-energies. However, for any practical calculation one needs certain
truncated approximate schemes. In the weak-coupling limit, one can restrict
the perturbation series for the Green's function to a certain order. Then, as
far as conservation laws are concerned, one encounters no particular problems,
as conserved quantities are conserved order by order in perturbation theory.
On the other hand, such perturbative expansion may not be adequate, as, for
example, in the strong coupling limit, where re-summation concepts have to be
applied, which re-sum certain sub-series of diagrams to any order. For such
re-summation schemes, the situation with conservation laws is not that
obvious.

We consider so-called $\Phi$-derivable approximations, first introduced by
Baym \cite{Baym} within the imaginary time method. Such approximations are
constructed by confining the infinite diagrammatic series for $\Phi$ either to
a set of a few diagrams or to some sub-series of diagrams.  Note that the
approximate $\Phi^{\scr{(appr.)}}$ itself is constructed in terms of ``full''
Green's functions and ``full'' classical fields, where ``full'' now implies
that we have to self-consistently solve the classical-field and Dyson
equations with the driving terms derived from this $\Phi^{\scr{(appr.)}}$
through relations (\ref{varphdl}) and (\ref{varphdl1}). It means that even
restricting ourselves to a single diagram in $\Phi^{\scr{(appr.)}}$, in fact,
we deal with a whole sub-series of diagrams in perturbation theory. Thereby,
the term ``full'' takes the sense of the sum of this whole sub-series. Thus, a
$\Phi$-derivable approximation offers a natural way of introducing closed, and
therefore self-consistent approximation schemes based on summation of
diagrammatic sub-series. In order to preserve the symmetry of the exact $\Phi$
with respect to permutations among $\ii \Lint(x_1) \dots \ii \Lint(x_n)$ (see
eq. (\ref{Phi-diagrm})), we postulate that a $\Phi$-derivable approximation
either takes into account all the diagrams up to a certain order $n$ (in terms
of ``full'' Green's functions and classical fields) or confines the treatment
to only those diagrams with topologically equivalent vertices. As a
consequence, approximate forms of $\Phi^{\scr{(appr.)}}$ define {\em
effective} theories, where $\Phi^{\scr{(appr.)}}$ serves as a generating
functional for the approximate source currents $\jba^{\scr{(appr.)}}(x)$ and
self-energies $\Sa^{\scr{(appr.)}}(x,y)$ (see eqs (\ref{varphdl}) and
(\ref{varphdl1}))
%
\begin{equation}\label{varphdl/appr}
\ii\jba^{\scr{(appr.)}}(x)
=\frac{\delta\ii \Phi^{\scr{(appr.)}}}{\delta 
\left(\Pba^{\scr{(appr.)}\ast} (x)\right)},
\end{equation}
\begin{equation}\label{varphdl1/appr}
-\ii
\Sa^{\scr{(appr.)}}(x,y)=  
\frac{\delta\ii \Phi^{\scr{(appr.)}}}{\delta \ii\Ga^{\scr{(appr.)}}(y,x)}
\times\left\{\begin{array}{ll}
2\quad&\mbox{for neutral fields,}\\
1\quad&\mbox{for charged fields,}
\end{array}\right.
\end{equation}
%
which then are the driving terms for the equations of motion for the classical
fields and propagators. The approximate $\Phi$ also provides the corresponding
expression for ${\cal E}^{\scr{int}}$ (see eq. (\ref{varphdl2})). Below, we
omit the superscript ``appr.''.

We now like to demonstrate that $\Phi$-derivable approximations possess a
number of remarkable properties. For such approximations, the invariances of
$\Phi$ play as central a role as the invariances of the Lagrangian for the
full theory.  Thereby, the variational principle, where the interaction
strength $\lambda(x)$, the classical fields $\Pba(x)$, and propagators
$\Ga(x,y)$ can be varied independently, provides a set of useful identities
and relations.

A general invariance of $\Phi$ is provided by the substitution $x\Rightarrow
x+\xi(x)$ for all integration variables in the contour integrations defining
$\Phi$. The Jakobi determinant required for each integration variable
can be accommodated by a simultaneous change of the scale function
$\lambda(x)$ at each vertex. Thus, the simultaneous variation
%
\begin{eqnarray}\label{x-xi}
\Pba(x)&\Rightarrow&\Pba(x+\xi(x)),
\nonumber\\
\Ga(x,y)&\Rightarrow&\Ga(x+\xi(x),y+\xi(y)), 
\\
\lambda(x)=1&\Rightarrow&\lambda(x)=\det\left(\delta^{\nu}_{\mu}
+\frac{\partial \xi_{\mu}}{\partial x_{\nu}}\right),
\quad\mbox{i.e.}\,\,\,
\delta\lambda(x)=\frac{\partial \xi_{\mu}}{\partial
x_{\mu}},\nonumber
\end{eqnarray}
leaves $\Phi$ invariant. This way, one deduces
%
\begin{eqnarray}\label{invar}
\hspace*{-4mm}\delta\Phi&=&
\suma\left\{
\oint\di x
\left[
\jba^*(x)\frac{\partial \Pba(x)}{\partial x_{\mu}}
      +\jba(x)\frac{\partial \Pba^*(x)}{\partial x_{\mu}}
\right] \xi_{\mu}(x)\right.
\nonumber
\\
\hspace*{-4mm}
&&\hspace*{1.9cm}\left.-\ii\oint\di x\di y\Sa(x,y)
\left[
\frac{\partial \Ga(y,x)}{\partial x_{\mu}}\xi_{\mu}(x)
+\frac{\partial \Ga(y,x)}{\partial y_{\mu}}\xi_{\mu}(y)
\right]\right\}
\nonumber\\
&&-\oint\di x{\cal E}^{\scr{int}}(x)\frac{\partial \xi_{\mu}}{\partial
x_{\mu}} =0.
\end{eqnarray}
%
Interchanging
$x$ and $y$ in the second $\Sa $ term in squared brackets, 
using partial integration and that the transformation $\xi(x)$ can be 
chosen arbitrarily, one 
obtains the following relation
%
\begin{equation}\label{keyeq}
\begin{array}{rcl}\displaystyle
\frac{\partial}{\partial x_{\mu}}{\cal E}^{\scr{int}}(x)
&+&
\suma
\left[
\jba^*(x)\frac{\partial \Pba(x)}{\partial x_{\mu}}
+\jba(x)\frac{\partial \Pba^*(x)}{\partial x_{\mu}}\right]
\\
&-&\displaystyle \ii\suma
\oint\di y\left[ \Sa(x,y)\frac{\partial \Ga(y,x)}{\partial x_{\mu}}+
\frac{\partial \Ga(x,y)}{\partial x_{\mu}}\Sa(y,x)\right] =0.
\end{array}
\label{epsilon-inv}
\end{equation}
%
This is the key relation to prove energy-momentum conservation. It has
features similar to a Ward identity, as it links derivatives of one-point
functions with those of two-point functions.  The two-point function
contribution to this expression is of type of eq.  (\ref{diffrules1}), so that
in eq.  (\ref{keyeq}) the differentiations of the discontinuities, indeed,
cancel out.

With the help of the equations of motion (\ref{eqmotion1}),
(\ref{Dyson}) and (\ref{DysonI}), the divergence of the kinetic term
of the energy-momentum tensor $\Theta^{\mu\nu}$ 
(\ref{E-M-G-tensor}) can be cast into
%
\begin{eqnarray}
\label{E-M-free-tensor-av-div}
\nonumber
&&
\frac{1}{2}\suma\partial_{\mu}\left[
\left(\left(\widehat{p}^{\nu}_x\right)^* 
+ \widehat{p}^{\nu}_y\right)
\left(\left(\widehat{p}^{\mu}_x\right)^* 
+ \widehat{p}^{\mu}_y\right)
\left(\vphantom{\frac{1}{2}}
\Pba^*(x)\Pba(y) +\ii\Ga^{\scr{sym}} (y,x)
\right)
\right]_{x=y} 
\nonumber
\\
&&\hspace*{2cm}
=\partial_{\mu} g^{\mu\nu} {\cal E}^{\scr{pot}}(x)
-\suma
\left\{
\vphantom{\oint}
-\left[ \jba(x)\partial_{\nu}\Pba^*(x)
+\jba^*(x)\partial_{\nu}\Pba(x)\right] 
\right.\nonumber
\\
&&\displaystyle\left.  \hspace*{2cm}+\ii  
\oint\di
z\left[\Sa(x,z)\cdot\partial_{\nu}^{x}\Ga(z,x)
+\partial_{\nu}^{x}\Ga(x,z)\cdot\Sa(z,x)\right]\right\}.
\end{eqnarray}
%
The remaining non-full-derivative terms on the r.h.s. of this equation are
undesired, since they prevent us from presenting of the entire equation in the
form of a full divergence. However, they can also be transformed into a full
derivative form by means of identity (\ref{keyeq}).  Adding this identity to
eq. (\ref{E-M-free-tensor-av-div}) cancels out the undesired last braced
terms. Gathering all surviving terms of (\ref{E-M-free-tensor-av-div}), we
recognize it as the energy-momentum conservation law $\partial_\mu
\Theta^{\mu\nu}(x)=0$ with the energy-momentum tensor given by the Noether
expression (\ref{E-M-G-tensor}).  Hence, the existence of a conserved
energy-momentum tensor is proven for any $\Phi$-derivable approximation.

Along similar lines charge conservation can be proven, assuming that $\Phi$ is
invariant under the following simultaneous variation of classical fields and
propagators 
%
\begin{equation}
\label{c-global-tr1}
\Pba(x)\Rightarrow e^{-\ii e\Lambda (x)}\Pba(x),\quad
\Pba^*(x)\Rightarrow e^{\ii e\Lambda (x)}\Pba^*(x),\quad
\Ga(x,y)\Rightarrow e^{-\ii e\Lambda (x)}
\Ga(x,y)e^{\ii e\Lambda (y)}. 
\end{equation}
%
Applying the rule (\ref{varphi'}) of the $\Phi$ variation, to linear order
in the phase $\Lambda$, one obtains
%
\begin{eqnarray}
\label{U(1)-var1}
\delta\Phi 
&=&  e
\oint\di x 
\left[\jba^*(x)\Pba(x)-\jba(x)\Pba^*(x)\right]
\left[- \ii \Lambda(x)\right]
\nonumber
\\
&+& \ii e\oint\di x\di y
\left[\Sa(x,y)\Ga(y,x)-
       \Ga(x,y)\Sa(y,x)\right]
\left[- \ii \Lambda(x)\right]. 
\end{eqnarray}
%
Here we have used the variation rules for complex fields. 
Equating $\delta\Phi = 0$ for arbitrary $\Lambda$, we arrive at the
identity 
%
\begin{equation}\label{invarJ}
e
\left[\jba^*(x)\Pba(x)-\jba(x)\Pba^*(x)\right]
{\displaystyle + \ii  e\oint\di y
\left[\Sa(x,y)\Ga(y,x)-\Ga(x,y)\Sa(y,x)\right]}=0. 
\end{equation}
%
Note that the integral term of this identity is independent of the contour
placement of the $x$ variable due to discontinuity relation (\ref{diffrules0})
and, therefore, it is only a function of the physical value of $x$.

By means of equations of motion (\ref{eqmotion1}), (\ref{Dyson}) and
(\ref{DysonI}), the divergence of the Noether current of
eq. (\ref{c-G-current}) is seen to vanish
%
\begin{eqnarray}
\label{current-phi-div}
\hspace*{-5mm}
\ii\partial_{\mu} j^{\mu}
= e
\left[\jba^*(x)\Pba(x)-\jba(x)\Pba^*(x)\right]
{\displaystyle + \ii e\oint\di y
\left[\Sa(x,y)\Ga(y,x)-
       \Ga(x,y)\Sa(y,x)\right]}=0,
\end{eqnarray}
%
according to eq. (\ref{invarJ}). Thus, we have arrived at the current
conservation for any $\Phi$-derivable approximation, which is invariant
under (\ref{c-global-tr1}).

Similarly, one may derive the relation, resulting from the Lorentz invariance
of the $\Phi$ functional, which permits to demonstrate the conservation of the
angular momentum. However, we do not consider it here, since the
angular-momentum conservation is not of practical use in kinetics.  

Further
invariances generally depend on the properties of the interaction vertices in
the theory considered. For instance, other invariances result from the
property of fixed number of field operators attached to the interaction
vertices in a theory. This property provides relations between densities of
the interaction ${\cal E}^{\scr{int}}(x)$ and potential ${\cal
  E}^{\scr{pot}}(x)$ energies. If the number $\gamma$ of field
operators per vertex is fixed, the functional $\Phi$ is invariant under the
simultaneous local scaling of the interaction vertices by $\lambda(x)$, and the
classical fields and propagators by $(\lambda(x))^{-1/\gamma}$ and
$\left(\lambda(x)\lambda(y)\right)^{-1/\gamma}$, respectively. This invariance
provides relation (\ref{gamma}) at the level of expectation values.

\section{Thermodynamic Consistency}
\label{Consistency}

In the thermal equilibrium the density matrix is explicitly known,
cf. \cite{Land80},
%
\begin{eqnarray}
\label{ro-eq}
\widehat{\rho}^{\scr{eq}} = 
\frac{\exp\left(-\beta\Hh\{\mu\}\right)}{Z}, 
\end{eqnarray}
%
where $\beta = 1/T$ is
the inverse temperature, and $Z$ is the partition
function which is directly related to the thermodynamical potential,
%
\begin{eqnarray}
\label{om-eq-Z}
\Omega  = - T \ln Z.
\end{eqnarray} 
%
Since we deal now with thermodynamics, we have introduced 
the chemical potential $\mu$ in the conventional way, i.e. 
by adding to the Hamiltonian the relevant term 
%
\begin{eqnarray}
\label{H-mu} 
\Hh \{\mu\} = \Hh 
- \int \di^3 x \mu \medhat{j}^{0}_{\scr{(Noether)}}(x),
\end{eqnarray}
%
where $\medhat{j}^{0}_{\scr{(Noether)}}$ is the time-component of the
charged current
 eq. (\ref{c-pi-current}), now with $e=1$. 

We can use the same trick as that in the Matsubara technique, i.e. use
the fact that the equilibrium density matrix formally coincides with
evolution operator in the imaginary time. In the definition of the
$W$ functional (\ref{W}) we explicitly write $\Tr
\widehat{\rho}^{\scr{eq}}...$ instead of $\left<...\right>$. Thus,
taking into account that $\Gamma=W$ at 
vanishing external sources, we arrive at the following form of
$\Gamma$ functional in equilibrium 
\newcommand{\eqoint}{\int_{\cal C_{\scr{eq}}}}
%
\begin{eqnarray}
\label{W-eq}
\hspace*{-5mm} 
\Gamma^{\scr{eq}}\left\{\phi,\phi^*,\Gr,\lambda,\mu \right\}
=-\ii\ln\left(\frac{1}{Z} \Tr 
\exp\left[-\ii\eqoint\di t \Hh_{\scr{I}}^{0}\{\mu\} \right]
\Tc\exp\left[ \ii\eqoint\di x
\lambda\Lint_{\scr{I}}\{\mu\} 
   \right]\right),
\end{eqnarray}
%
with the integration contour ${\cal C_{\scr{eq}}}$ now being the 
sum of the real-time Schwinger-Keldysh contour (see figure 1) and the
imaginary-time Matsubara 
contour, i.e. it starts from an initial time $t_0$ goes to infinity,
then back to this initial time and after that, to $t_0-\ii\beta$. 
Taking into account the fact that $\Gamma = 0$ for the physical values
of $\phi$, $\phi^*$, $\Gr$, and $x$-independent $\lambda$, we obtain 
for the value of the thermodynamic potential 
(\ref{om-eq-Z}) 
%
\begin{eqnarray}
\label{Om-eq}
\hspace*{-4mm} 
\Omega \{\phi,\phi^*,\Gr,\lambda,\mu\}
= - T \ln \left\{\Tr 
\left(
\exp\left[-\ii\eqoint\di t \Hh_{\scr{I}}^{0}\{\mu\} \right]
\Tc\exp\left[ \ii\eqoint\di x
\lambda\Lint_{\scr{I}}\{\mu\} 
   \right]\right)\right\},
\end{eqnarray}
%
where the integral over the real-time section of the contour gives
zero. Hence, in eq. (\ref{Om-eq}) we can make the
replacement 
%
\begin{eqnarray}
\label{C_FS->C_M}
\eqoint\di t ... = \int_0^{-\ii \beta} \di t ... \quad . 
\end{eqnarray}
%
Thus, we have arrived at the proper thermodynamic representation of the
thermodynamic potential originally proposed by Luttinger and Ward
\cite{Luttinger}. Indeed, since all quantities under the integral are
analytically continued from the Schwinger--Keldysh contour to the Matsubara
contour, $\Omega$ is determined by the same expression as the
$\Gamma$ functional (\ref{PHIfunct}) but in terms of the Matsubara Green's
functions with the thermodynamic $\Phi_T$ functional represented by the same
set of closed diagrams. Thus, in the momentum representation from eq.
(\ref{Om-eq}) we arrive at
%
\begin{eqnarray}
\label{Omlat}
\Omega \left\{\phi,\phi^*,\Gr,\lambda,\mu \right\}
&=&
- \int \di^3 x \Lg^0\{\phi,\partial_\mu\phi\}
+T \suma \sum_{\omega_n} 
\int \di^3 x \frac{\di^3 p}{(2\pi)^3}
\exp(\ii\omega_n \eta )
\nonumber \\
&& \times 
\left( -\ln[-\Ga (\omega_n ,\vec p )] 
+ 
\vphantom{\int}
\Sa(\omega_n ,\vec p )\Ga (\omega_n ,\vec p )\right)+\Phi_T , \quad\eta
\rightarrow 0 , 
\end{eqnarray}
%
where $\Phi_T=-\ii T\Phi$, $\omega_n =2\pi \ii nT $, and summation runs over
Matsubara frequencies. In the standard way (see, e.g., ref. \cite{Abrikos}) by
converting the $\omega_n$-sum in eq. (\ref{Omlat}) into the energy integral
expressed in terms of the real-time quantities of eqs (\ref{Geq}) and
(\ref{Seq}), this thermodynamic potential is also easily expressed in terms of
the real-time quantities
%
\begin{eqnarray}
\label{Om-maz-cont} 
\Omega \left\{\phi,\phi^*,\Gr,\lambda,\mu \right\} &=&
- \int \di^3 x \Lg^0\{\phi,\partial_\mu\phi\}
+\suma \int \di^3 x \dpi{p} n({\varepsilon})
\nonumber
\\ 
&\times&
\left(-2\mbox{Im} \; 
\ln\left[-
\Ga^R(\varepsilon +\ii 0 ,\vec p )
\right]
\vphantom{\int}
-
\vphantom{\int}
 \Re\Ga^R\Gm -\A\Re\Sa^R\right)
+\mbox{\large $\Phi_T$},\quad
\end{eqnarray} 
%
where 
%
\begin{equation}\label{occup}
n(\varepsilon) =\left[\exp(\varepsilon/T)- 1\right]^{-1}\;  
\end{equation}
%
is the thermal Bose--Einstein occupation number.  Thus, the problem of the
thermodynamic consistency can immediately be re-addressed from the
Schwinger--Keldysh approach to the Matsubara one. Within the Matsubara
formalism, this problem was considered by Baym \cite{Baym}. He has shown that
any $\Phi$-derivable approximation to the thermodynamic potential is
thermodynamically consistent. Hence, we have proved that our $\Phi$ derivable
approximations to the $\Gamma$-functional are also thermodynamically
consistent.

The stationary property of the $\Gamma$ functional (and, hence, of $\Omega$)
with respect to variations in full Green's functions and classical fields,
eq. (\ref{varG/phi}), is the key feature of $\Gamma$ that provides the
thermodynamic consistency. It implies that any derivative of the thermodynamic
potential to any thermodynamic parameter like $\beta$ or $\mu$ is then given
by accounting only the explicit dependence of $\Omega$ on these parameters,
since the implicit dependences through $\Ga$ and $\phi$ drop out due to the
stationarity property. Therefore $\Phi$-derivable approximations preserve the
corresponding thermodynamic relations as for the exact partition sum, and thus
provide thermodynamic consistency.

\section{Conclusion}

Our aim was to develop a regular way of constructing self-consistent
approximations to quantum transport theory within the formalism of
non-equilibrium Green's functions on the real-time contour, developed
by Schwinger, Kadanoff, Baym and Keldysh \cite{Schw,Kad62,Keld64}. 

We have employed functional methods for Green's functions on the real-time
contour. Our main result is the definition of a generating functional $\Phi$
which is determined by the sum of all closed (i.e. without external points)
skeleton diagrams in terms of classical fields and full Green's functions on
the real-time contour. The main feature of this $\Phi$ functional is that it
plays a similar role as the interaction Lagrangian but in the space of
classical fields and full Green's functions on the contour. This means that
all important quantities of a system (such as sources of classical fields,
self-energies, interaction energy, etc.)  can be derived by respective
variations of the $\Phi$ functional.  Within thermodynamics, the $\Phi$
functional was introduced by Luttinger and Ward \cite{Luttinger} and later
used by Baym \cite{Baym}. Our treatment extends the definition of the $\Phi$
functional to any non-equilibrium system. Following the thermodynamic
treatment of Cornwell et al. \cite{Cornwell}, we have also extended the
non-equilibrium $\Phi$ functional to the case of non-vanishing classical
bosonic fields. This last generalization allows us to self-consistently
describe the dynamics of both the order parameter (the classical field) and
fluctuations on equal footing, e.g. in the theory of phase-transition
phenomena.

The advantage of the $\Phi$ functional is that we may formulate various
approximations at the level of $\Phi$, thus defining so called
$\Phi$-derivable approximations. In particular, we may construct effective
theories right at the level of Green's functions and effective vertices. These
approximations possess some important features: they respect exact
conservation laws on the level of expectation values (with the Noether values
for the conserved quantities) and have a proper thermodynamic limit. Note that
other approximation schemes, e.g. at the level of self-energies, far not
always possess such properties.

The question of consistency becomes especially important for a multi-component
system, where the properties of one species can change due to the presence of
interaction with the others and {\em vice versa}. The ''{\em vice versa}'' is
very important and corresponds to the principle of {\em actio} = {\em
re-actio}. This implies that the self-energy of one species cannot be changed
through the interaction with other species without affecting the self-energies
of the latter ones also. The $\Phi$-derivable scheme offers a natural and
consistent way to account for this principle. Within thermodynamic
considerations this has recently been considered for the interacting pion -
nucleon -delta-resonance system, where the coupling to the delta resonance
leads to a softening of the pion modes below the resonance mass
\cite{Weinhold} and for a relativistic QED plasma in \cite{Baym98}.

For the relativistic scheme considered here we argue that a careful
construction of conserved quantities requires symmetric expressions in terms
of $\Gr^{-+}$ and $\Gr^{+-}$ Green's functions ($\Gr^{<}$ and $\Gr^{>}$ in the
Kadanoff--Baym notation, respectively). This is in contrast to expressions
involving only $\Gr^{-+}$ Green's function, which are often used in the
literature. These symmetric expressions describe contributions of both
particles and anti-particles on equal footing, as well as take a proper
account of modification of the vacuum polarization in the medium. Of course,
these symmetric expressions still require a proper vacuum renormalization to
be done in any actual calculation.

We have done our consideration on the example of relativistic bosons, since
they allow us to demonstrate the basic features of this approach --- inclusion
of both Green's functions and classical fields into the scheme --- without
extra complications. The generalization to multi-component systems is
straightforward. Here, we expect a consistent description of chiral $\sigma$-,
$\pi$- condensates together with fluctuations, as an immediate
application. The generalization to vector mesons and/or fermions, though more
technically involved due to the resulting tensor structure of the propagators
and self energies is also straight forward. The formalism can also be
generalized to theories with derivative couplings in the interaction
Lagrangians. While the expressions for the equations of motion and the
energy-momentum tensor receive additional terms from the derivative couplings,
the diagrammatic rules are the same as given here.

The developed scheme of constructing self-consistent approximations provides a
suitable basis for the derivation of kinetic equations beyond the limitations
of the quasi-particle approximation. Such generalized transport schemes
respect parts of the quantum nature of the particles and, in particular, take
account of their finite mass-widths. The finite mass-width may be either
inherent in a particle already from its vacuum properties (e.g., resonances)
or may be acquired by a stable particle in a dense environment due to frequent
interactions. In the case of nuclear collisions at intermediate ($\sim$ 1
GeV/nucleon) to ultra-relativistic energies, one encounters mean
single-particle energies in the range of the typical temperature of $T=50$ -
$200$ MeV. Important resonances, like the delta-resonance or the rho-meson,
have decay widths beyond $\sim$ 100 MeV, while typical collision rates
estimated from presently used quasi-particle transport schemes are typically
in the order of $T$. These circumstances definitely prevent quasi-particle
based transport codes from providing reliable results for such collisions. The
main steps in the derivation of self-consistent and tractable transport
equations for particles with finite width will be published in a forthcoming
paper \cite{IKV}.

\section*{Acknowledgments}
We are grateful to G.E. Brown, P. Danielewicz, B. Friman, H. van Hees, and
E.E. Kolomeitsev for fruitful discussions and suggestions.  Two of us
(Y.B.I. and D.N.V.) highly appreciate the hospitality and support afforded to
them at Gesellschaft f\"ur Schwerionenforschung.  This work has been supported
in part by BMBF under the program on scientific-technological collaboration
(WTZ project RUS-656-96).\\[3mm] 

\appendix{APPENDICES} 
\section{Diagram Rules for $\Phi$, $J$, and $\Pi$}\label{Diag-rules}

The interaction vertex function $V(x)$ entering the diagram is normalized in
the standard way, c.f. \cite{Itz80}, i.e. with factors $n!$ relative to
$\Lint(x)$ for each type of operator occuring with multiplicity $n$ in the
vertex. E.g., the vertex function simply becomes $- \ii V(x_k)=-\ii g$ for
$\Lint = -g \phi^{4}/4!$ (4 identical operators) and for $\Lint = -g
(\phi^{\ast}\phi)^{2}/(2!\cdot 2!)$ (twice two identical operators).  The
diagrammatic rules to calculate $\ii\Phi$, $\ii J(x)$ and $-\ii\Pi$ for a
given theory are as follows
\begin{itemize}
\item[(i)] Draw all topologically distinct, closed and entirely
connected diagrams with $N$ internal vertices $x_1, x_2, ...
x_{N}$, where classical field pins and propagator lines
saturate the valences of all vertices in the diagram,
c.f. (\ref{phi4diagrams}) above. Closed diagrams for $\ii\Phi$ have no 
external points, while  $\ii J(x)$ has one external point, and  
$-\ii\Pi$ has two external points. For charged bosons, pins and
propagator lines have an arrows sense,  distinguishing $\Pa$ from 
$\Pad$ at the vertices, the sense direction pointing towards $\Pa$. 
\item[(ii)] For $\ii\Phi$, $\ii J(x)$ and $-\ii {\cal E}^{\scr{int}}(x)$
keep only those diagrams that are two-particle
irreducible, i.e. which cannot be split into two pieces by cutting two
different propagator lines. For $-\ii\Pi$ keep only those diagrams which
result from $\Phi$ by opening one propagator line.
\item[(iii)] To each line, connecting $x_l\longrightarrow x_k$, assign
the factor $\ii \Ga (x_k,x_l)$. 
\item[(iv)] To each pin attached to $x_k$, assign the factor
$\Pba(x_k)$ or $\Pba^*(x_k)$ depending on the sense. 
\item[(v)] To each vertex $x_k$ assign the vertex factor $- \ii
V(x_k)$ as determined by $\Lint(x)$. 
\item[(vi)] Integrate all internal $x_1, x_2, ... x_{N}$
over the contour. 
\item[(vii)] Multiply the result by the symmetry factor $S$, which is
calculated as follows\\ $1/N_G!$ factor for every $N_G$ equivalent internal
lines, \\ $1/N_{\phi}!$ factor for every $N_{\phi}$ classical fields entering
each vertex, \\ $1/2$ factor for every self-closed line loop (tad-poles) for
{\em real} fields.
\item[(viii)] Sum all diagrams. For $\ii\Phi$ (contrary to $\ii J(x)$ and 
$-\ii\Pi$), put the extra factor $1/n_\lambda$ for each diagram, where
$n_\lambda$ counts the number of vertices in the diagram.
\end{itemize}

In many cases like in transport treatments,  it
is advantages to consider the diagrams decomposed into the two contour
sections at each vertex, e.g., to calculate quantities like
$\Sb^{-+}$ and $\Sb^{+-}$ self-energies in terms of exact Green's
functions. 
Therefore, ''physical''-time diagrammatic rules in
the matrix scheme are also required. Here we present only those rules which
differ from the above ones on the real-time contour, bearing in mind that all
other rules remain valid:
\begin{itemize}
\item[(iii$'$)] To each internal vertex $x_k$ first assign a sign
$i_k\in\{+,-\}$ defining the contour placement $x_k^{i_k}$.  To each line,
connecting $x_l^{i_l}\longrightarrow x_k^{i_k}$, assign the factor $\ii
\Ga^{i_ki_l} (x_k,x_l)$, $i_k,i_l \in \{+,-\}$.
\item[(vi$'$)] For all internal points integrate all $x_1, x_2, ... x_{N}$
over the real-time axis and space, for each internal $"+"$ vertex multiply by
$(-1)$ and finally sum over all internal contour placements $i_1, i_2,
... i_{N}$ ($i_k \in \{+,-\}$).
\end{itemize}

\section{Equilibrium Relations}\label{eq.rel}

For completeness of the thermodynamic consideration, we explicitly
present here equilibrium relations between quantities on the real-time
contour. Basically, they follow from the Kubo--Martin--Schwinger
condition \cite{Kubo} 
%
\begin{eqnarray}
\label{KMS-G}
\Ga^{-+}(p) = \Ga^{+-}(p) e^{-\varepsilon/T}, \,\,\,\,
%
%
\Sa^{-+}(p) = \Sa^{+-}(p) e^{-\varepsilon/T}, 
\end{eqnarray}
%
where $\varepsilon = p_\nu U^\nu - \mu$ with $U^\nu$ and $\mu$ being a
global 4-velocity of the system and a chemical potential related to
the charge, respectively. All the Green's functions can be 
expressed through retarded or advanced Green's functions: 
%
\begin{eqnarray}
\label{Geq}
\left( \Ga^{i,j} (p) \right)= 
\left(\begin{array}{ccc}       
\left[1+ n(\varepsilon)\right]\Ga^R(p)- n(\varepsilon) \Ga^A(p) &&
-\ii n(\varepsilon) \A(p)\\[3mm]  
-\ii \left[1+ n(\varepsilon)\right] \A(p) &&
-\left[1+ n(\varepsilon)\right] \Ga^A(p)+ n(\varepsilon) \Ga^R(p)
\end{array}\right), 
\end{eqnarray}
%
$i,j \in \{+,-\}$, and the self-energies take a similar form 
%
\begin{eqnarray}
\label{Seq}
\left( \Sa^{i,j} (p) \right) = 
\left(\begin{array}{ccc}       
\Sa^R(p) - \ii n(\varepsilon) \Gm(p) && 
-\ii n(\varepsilon) \Gm(p) \\[3mm]
-\ii \left[1+ n(\varepsilon)\right]\Gm(p)&&
- \Sa^A(p) - \ii n(\varepsilon) \Gm(p) 
\end{array}\right). 
\end{eqnarray}
%
Here $n(\varepsilon)$ is the thermal Bose--Einstein occupation number
defined in eq. (\ref{occup}). 


\end{document}